\newif\ifcheckpagelimits
\checkpagelimitstrue
\checkpagelimitsfalse

\ifcheckpagelimits
 \documentclass[nofootinbib,pra,aps,twocolumn,showpacs,showkeys,%
 amsmath,amssymb,superscriptaddress,final,reprint,floatfix,longbibliography]{revtex4-1}
 \newcommand{\todo}[1]{}
\else
 \documentclass[pra,aps,twocolumn,showpacs,showkeys,%
 amsmath,amssymb,superscriptaddress,final,reprint,floatfix,longbibliography]{revtex4-1}
 \newcommand{\todo}[1]{{\pdfmargincomment[icon=Note,color=pink]{#1}}}
\fi

\usepackage[table]{xcolor}
\usepackage{helvet} 
\usepackage{lineno}
  \usepackage{mathptmx}
\usepackage{subfigure}
\usepackage{hhline} 
\usepackage{psfrag,graphicx}
\usepackage{dcolumn}
\usepackage{amsmath,amssymb}
\usepackage{bm}
\usepackage{color}
\usepackage{overpic}
\usepackage{latexsym}
\usepackage{epstopdf}
\usepackage{color}
\usepackage[english]{babel}
\usepackage{latexsym}
\usepackage{stmaryrd}

\usepackage{braket}

\definecolor{mygrey}{gray}{0.94}
\definecolor{myblue}{rgb}{0.2,0.2,0.8}
\definecolor{myzard}{cmyk}{0,0,0.05,0}
\definecolor{mywhite}{rgb}{1,1,1}
\definecolor{myred}{rgb}{1,0.,0.3}

\usepackage[colorlinks=true,citecolor=myblue,linkcolor=myred]{hyperref}
\DeclareMathAlphabet{\mathpzc}{OT1}{pzc}{m}{it}

 \def\ee{\mathord{\rm e}}
 
 \def\ii{\mathord{\rm i}}
\def\min{\mathord{\rm min}}

\def\half{\textstyle\frac{1}{2}}

\renewcommand{\ii}{{\rm i}}
\renewcommand{\ee}{{\rm e}}
\def\beq{\begin{equation}}
\def\eeq{\end{equation}}

\def\barray{\begin{eqnarray}}
\def\earray{\end{eqnarray}}

\definecolor{LightGrey}{gray}{0.85}
\definecolor{LightGreen}{rgb}{0.60,1,0.2}
\definecolor{UberLightGreen}{rgb}{0.80,1,0.6}
\definecolor{LightOrange}{rgb}{1,1,0.39}

\newlength\figureheight
\newlength\figurewidth

\def\ket#1{\left|#1\right>}
\def\bra#1{\left<#1\right|}

 \def\ee{\mathord{\rm e}}
 
 \def\ii{\mathord{\rm i}}
\def\min{\mathord{\rm min}}

\def\half{\textstyle\frac{1}{2}}

\renewcommand{\ii}{{\rm i}}
\renewcommand{\ee}{{\rm e}}

\definecolor{mygrey}{gray}{0.94}
\definecolor{myred}{rgb}{1,0.,0.3}

\begin{document}

\title{Fault-tolerant protection of near-term trapped-ion topological qubits under realistic noise sources}

\author{A. Bermudez}
\affiliation{Departamento de Fisica Te\'orica, Universidad Complutense, 28040 Madrid, Spain}

\author{X. Xu}
\affiliation{Department of Materials, University of Oxford, Parks Road, Oxford OX1 3PH, United Kingdom}

\author{M. Guti\'errez}
\affiliation{Department of Physics, College of Science, Swansea University, Singleton Park, Swansea SA2 8PP, United Kingdom}
\affiliation{Escuela de Qu\'imica, Universidad de Costa Rica, San Jos\'e, 2060 Costa Rica}
\author{S. C. Benjamin}
\affiliation{Department of Materials, University of Oxford, Parks Road, Oxford OX1 3PH, United Kingdom}

\author{M.~M\"uller}
\affiliation{Department of Physics, College of Science, Swansea University, Singleton Park, Swansea SA2 8PP, United Kingdom}

\begin{abstract}
The quest of demonstrating  beneficial quantum error correction in  near-term noisy quantum processors can benefit enormously from a low-resource optimization of fault-tolerant schemes, which are specially designed for a particular platform considering both state-of-the-art technological capabilities and main sources of noise. In this work, we show that flag-qubit-based fault-tolerant techniques for active error detection and correction, as well as for encoding of logical qubits, can be leveraged in current designs of trapped-ion quantum processors to achieve this break-even point of beneficial quantum error correction. Our improved description of the relevant sources of noise, together with detailed schedules for the implementation of these flag-based protocols, provide one of the most complete microscopic characterizations of a fault-tolerant quantum processor to date. By extensive numerical simulations, we provide a comparative study of flag- and cat-based approaches to quantum error correction, and show that the superior performance of the former can become a landmark in the success of near-term quantum computing with noisy trapped-ion devices.  
\end{abstract}


\maketitle

\makeatletter
\makeatother
\begingroup
\hypersetup{linkcolor=black}
\tableofcontents
\endgroup

\section{\bf Towards fault-tolerant (FT) quantum error correction (QEC) with trapped ions}
\label{sec:QEC_color_code}

\subsection{Introduction}
\label{sec:intro}

\subsubsection{Development and assessment of near-term QEC devices}

The prospect of processing information quantum-mechanically and, thereby, solving computational problems beyond the reach of classical devices~\cite{nielsen-book} has stimulated enormous research efforts, which aim at building and scaling up prototype quantum processors \cite{qc_review}. To date, the field of quantum computing has reached a considerable level of maturity, allowing for high-accuracy control over ever-larger qubit registers. These advances are expected to enable  the construction and operation of near-term devices estimated to contain about 30 to 100 qubits, which, for the first time, demonstrate quantum advantages \cite{Preskill2018}.

Currently, large efforts are focusing on identifying specific applications and algorithms, such as hybrid quantum-classical approaches \cite{Farhi-2014, EVS-2014, Hybrid-Q-Cl-2016,qve_chemistry,qve_schwinger}, which can be directly executed as low-depth  quantum circuits  on  available registers of bare  physical qubits.  In this way, one may demonstrate quantum advantage using faulty qubits without  the resource overhead required for quantum error correction (QEC)~\cite{qec_review}. On a  longer-term perspective, however, the construction of general-purpose large-scale fault-tolerant (FT) quantum processors will require encoding of information in logical qubits, and repetitive application of active QEC cycles to detect and correct errors occurring during storage and processing \cite{qec_shor,calderbank-pra-54-1098,steane-prl-77-793, FTQC}. Various physical platforms have emerged as promising candidate systems to build such scalable devices, including trapped ions~\cite{trapped_ion_qc}, and Rydberg atoms in optical lattices or trap arrays~\cite{rydberg_qc}, as well as solid-state platforms such as superconducting circuits~\cite{sc_qubits_qc}, nitrogen-vacancy centres~\cite{nv_qc}, or quantum dots~\cite{qdots_qc}.

A first generation of proof-of-principle implementations have demonstrated basic QEC codes in a variety of platforms
\cite{3_qubit_code_nmr, 5_qubit_code_nmr, 3_qubit_code_nist_ions, 3_qubit_code_superconducting, 3_qubit_code_repetitive_qec_ions, repetitive_QEC_NV}, including minimal versions of the topological color code \cite{nigg-science-345-302,phase_opt_color_code} and surface code \cite{surface_code_superconducting_line,surface_code_superconducting_square,stab_measurement_sc}. Current  efforts focus on the demonstration of fault-tolerance in near-term and potentially-scalable devices \cite{ gottesman_small_codes, ft_error_detection_monroe,ft_state_preparation_sc}, and the implementation of full QEC cycles on logical qubits in the parameter regime where they outperform their constituent physical qubits~\cite{eQual_1qubit}. On the theory side, an essential contribution to push these developments concerns {\it (i)} the development and optimisation of resource-efficient and fault-tolerant protocols especially designed for a particular platform, and {\it (ii)} the faithful modeling of the underlying quantum hardware and experimental noise processes, which is crucial to assess the performance of the first-generation  low-distance QEC codes. In this context, a series of studies have  shown that it is important to include realistic error sources such as non-Pauli errors~\cite{poulin_non_pauli,pauli_approx}, qubit leakage~\cite{fowler_leakage,suchara_leakage} and losses~\cite{barrett_loss,vodola_loss}, as well as spatially and temporally correlated noise~\cite{correlatd_noise}. Oversimplified single- or few-parameter noise models that neglect these  effects can lead to a drastic over- or under-estimation of the QEC prospects. 

In this work, we present a detailed theoretical study that aims at identifying the requirements and parameter regimes for beneficial QEC in state-of-the-art and near-future trapped-ion quantum processors. In our work, we focus on a thorough and realistic modeling of the experimental QEC toolbox in high-optical-access segmented ion traps~\cite{HOA2_trap}, which allow one to manipulate dual-species ion crystals~\cite{mixed_species_review} under cryogenic conditions. This architecture is scalable, as 1D trapping zones can be coupled via junctions into larger potentially 2D trap array structures forming a so-called quantum charge-coupled device (QCCD)~\cite{qccd}, which is complementary to approaches based on optical coupling of ions in separated traps~\cite{musiq}. We carefully model physical qubits as multi-level ions, with associated amplitude damping and leakage processes due to spontaneous decay of the electronic states. The experimentally available toolset, most prominently the set of single- and two-qubit entangling operations, are described with  non-Pauli or correlated error channels, as derived from a detailed and quantitative quantum optical modeling of the underlying microscopic electronic and vibrational dynamics.

On the QEC side, we choose to work with the distance-3 topological 7-qubit color code~\cite{bombin-prl-97-180501}. Here, we pay particular attention to recently proposed FT stabiliser readout protocols based on so-called flag qubits~\cite{flag_based_readout}, and compare their performance with other established FT readouts that involve a larger number of ancilla qubits~\cite{shor_ft_qec,aliferis_ft_qec}. For the flag-based syndrome measurement, we provide optimized and resource-efficient compilations of the required circuits into the trapped-ion gate primitives, and furthermore complement these QEC protocols for the correction of standard computational errors (bit and/or phase flips) by a new leakage suppression technique. Extensive Monte-Carlo wave function simulations of the QEC protocols and noise processes allow us to identify the parameter regimes in which reaching the break-even-point of useful QEC is expected to become feasible in realistic near-term trapped-ion based quantum information processors.

\subsubsection{Topological QEC and color codes}
As advanced above, a promising route for the extensibility of  prototype  quantum processors towards  large-scale FT quantum computers is the use of active QEC. Here, the  logical  information is redundantly encoded in several entangled data qubits  defining the so-called code subspace, such that errors drive the system out of this subspace, and can be detected and corrected  by measuring collective observables without damaging the encoded  information~\cite{qec_review}. A particularly-promising type of encoding is that offered by  {\it topological planar codes}, such as the surface version~\cite{surface_code} of Kitaev's toric code~\cite{toric_code} and topological color codes~\cite{bombin-prl-97-180501}. For both families of codes, the physical qubits can be arranged on a planar lattice, and the collective observables can be defined as local stabilizers~\cite{stabiliser}, i.e.multi-qubit Pauli operators that involve only groups of spatially neighboring qubits on the lattice. This locality of the check operators implies that only local quantum processing is required to detect and correct the possible errors. Besides this locality property, topological codes are particularly interesting due to the high FT threshold values~\cite{dennis-j-mat-phys-43-4452,raussendorf-prl-98-190504,threshold_color_code}, i.e.~reliable computations of arbitrary length will become feasible if the error per operation is below a certain threshold. 

\begin{figure}[t]
 \begin{centering}
  \includegraphics[width=1\columnwidth]{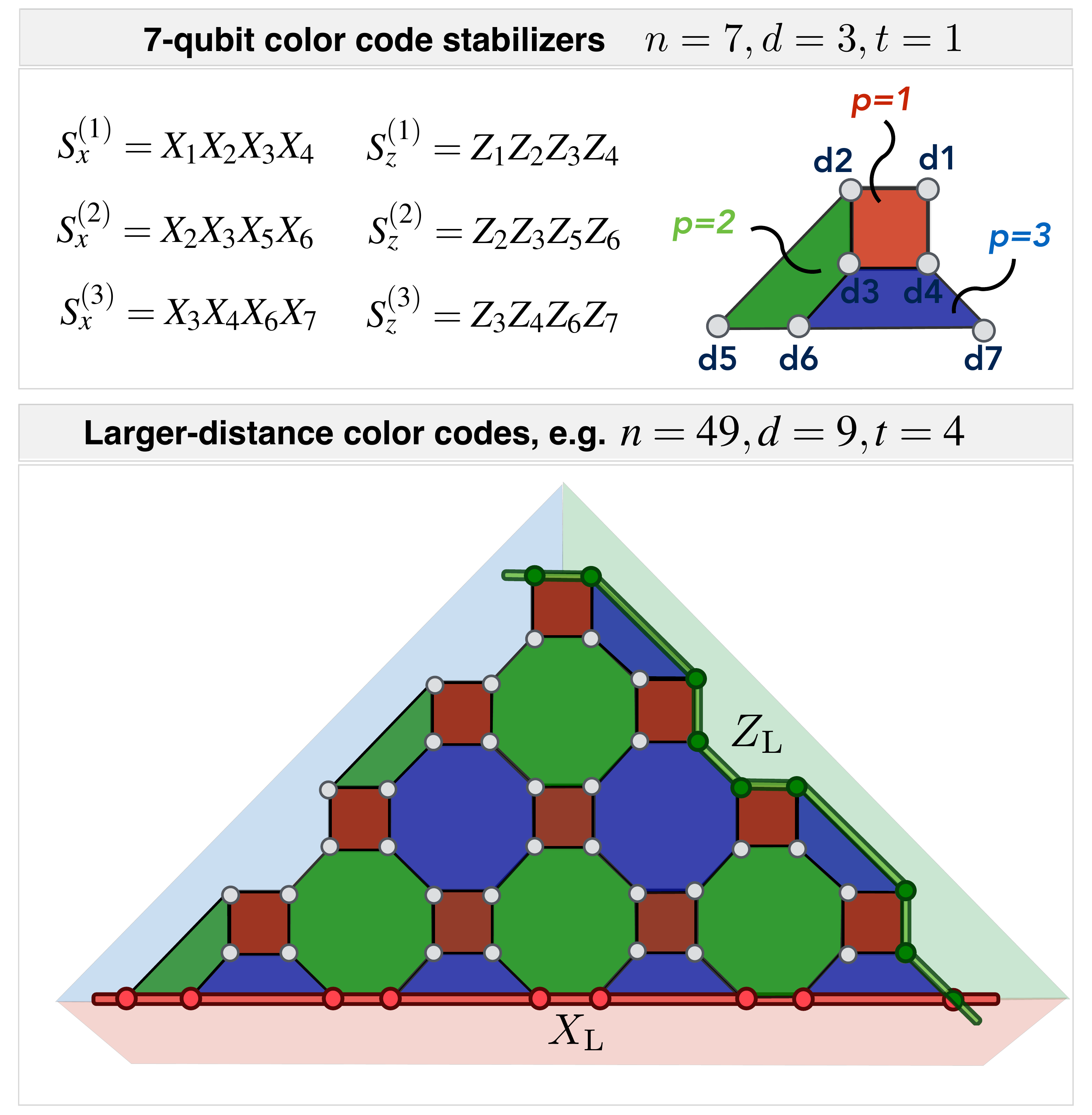}\\
  \caption{\label{Fig:7qubitCode} {\bf Color code scheme:}  (Upper panel) The quantum information is redundantly encoded in  $n=7$ data qubits forming a   planar code  with $n_{\rm p}=(n-1)/2=3$ plaquettes, leading to $s=2n_{\rm p}=6$ stabilisers, which yields $k=n-s=1$ logical qubits. The code space is defined via $S_x^{(p)}, S_z^{(p)}$ stabilizer operators, each acting on a plaquette  $p=1,2,3$ that involves four data qubits. The number of qubits along the  boundaries determine the distance of the code $d=3$, such that $t=(d-1)/2=1$ errors can be corrected. We note that the form of the plaquettes is a mere visualization, and they could be deformed such that an equilateral triangle is formed, as used in~\cite{nigg-science-345-302,eQual_1qubit}. (Lower panel) Larger color codes are constructed by growing a so-called 4.8.8.triangular lattice, which is a three-colorable tilling of the plane  with $n=(d^2+2d-1)/2$ data qubits, each of which belongs (in the bulk) to one square and two neighboring octagons, e.g. $d=9$, $n=49$. The logical operators, which can be defined in a bit-wise manner, can also be deformed into $X$- and $Z$-type colored strings  $X_{\rm L}$- and $Z_{\rm L}$ connecting two boundaries of a different color. Thereby, logical information is encoded globally, such that the local errors occurring at low enough rates on physical qubits have a smaller impact as the lattice size grows.}
\end{centering}
\end{figure}

We will present below a generic  trapped-ion toolbox  that can be used to implement any topological stabilizer code. However, one of our main goals is to understand the minimal requirements to prove the beneficial nature of  QEC in near-term trapped-ion experiments,  lying a set of building blocks to construct  future QEC experiments with ever-increasing registers. Therefore, we will here optimize the resources for the smallest, yet fully-functional $7$-qubit topological color code. This code, unitarily equivalent to Steane's code~\cite{steane_code}, is an instance of the so-called {\it triangular color codes} (see Fig.~\ref{Fig:7qubitCode})~\cite{bombin-prl-97-180501}, and allows one to store and manipulate a single logical qubit  redundantly encoded into  $7$ physical qubits. The code, which belongs to the family of CSS codes~\cite{calderbank-pra-54-1098,steane-prl-77-793}, can correct  a single error  due to either  bit   or  phase flips. The most-likely bit- and phase-flip errors can be inferred from the measurements of three $Z$- and $X$-type plaquette stabilizers, as shown in Fig.~\ref{Fig:7qubitCode},
\beq
\label{eq:stabilizers}
\begin{split}
&S^{(1)}_x=X_1X_2X_3X_4,\hspace{1ex}S^{(2)}_x=X_2X_3X_5X_6,\hspace{1ex}S^{(3)}_x=X_3X_4X_6X_7,\\
&S^{(1)}_z=Z_1Z_2Z_3Z_4,\hspace{1ex}S^{(2)}_z=Z_2Z_3Z_5Z_6,\hspace{1ex}S^{(3)}_z=Z_3Z_4Z_6Z_7.\\
\end{split}
\eeq
These commuting operators define the so-called code space  $\mathcal{V}_{\rm code}\subset \mathcal{H}$ spanned by all stabilizer eigenstates $\ket {\Psi}$ of eigenvalue +1, $S_\alpha^{(p)}\ket{\Psi}=\ket{\Psi}$. In this case, the code subspace is two-dimensional, encoding a single logical qubit $Z_{\rm L}\ket{0}_{\rm L}=\ket{0}_{\rm L}$, $Z_{\rm L}\ket{1}_{\rm L}=-\ket{1}_{\rm L}$, and $X_{\rm L}\ket{0}_{\rm L}=\ket{1}_{\rm L}, X_{\rm L}\ket{1}_{\rm L}=\ket{0}_{\rm L}$,  where $Z_{\rm L}=\otimes_iZ_i$, $X_{\rm L}=\otimes_iX_i$ are possible representations of the generators of the logical qubit.

\begin{figure}[t]
 \begin{centering}
  \includegraphics[width=1.0\columnwidth]{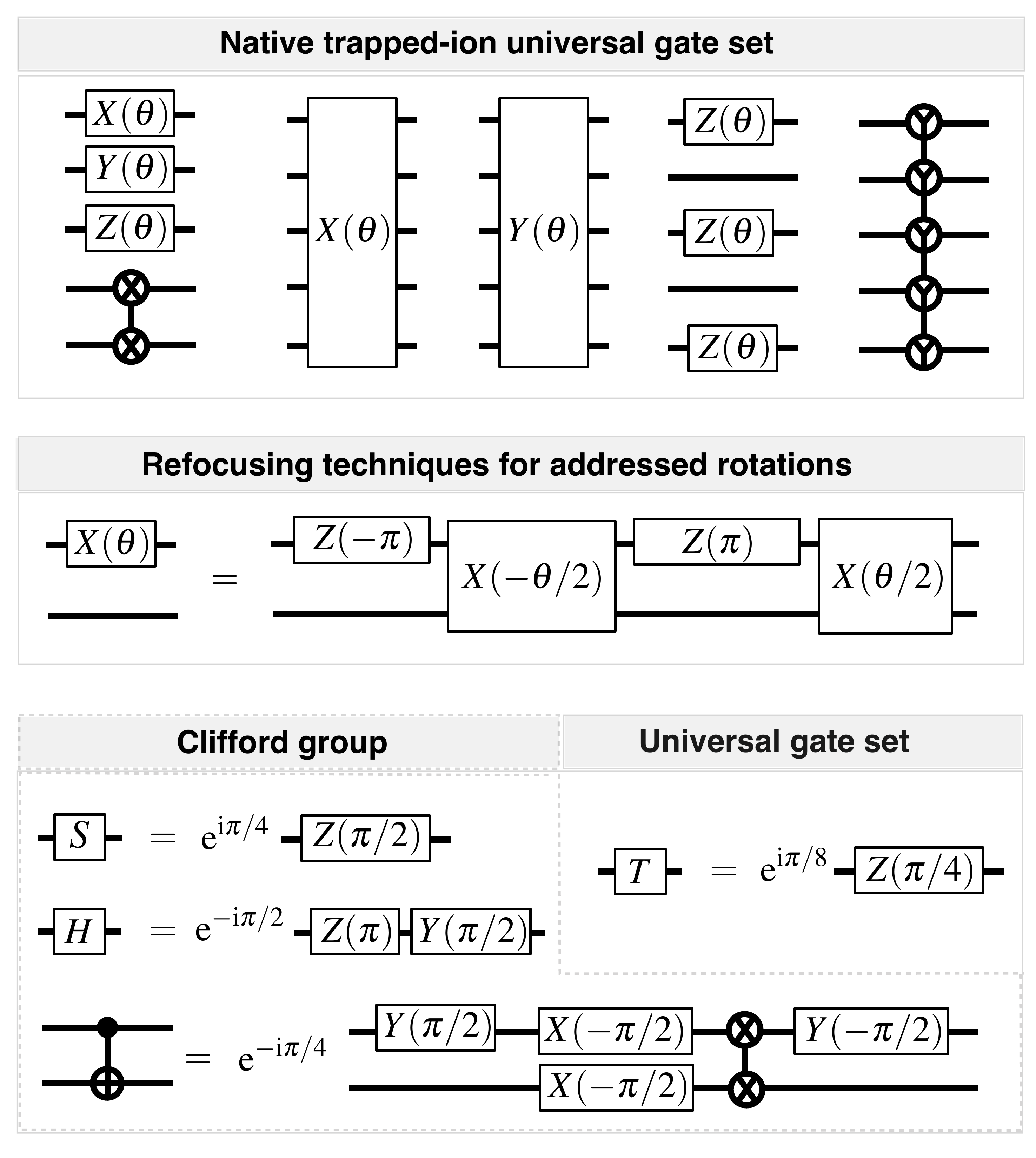}\\
  \caption{\label{Fig:gate_set} {\bf Trapped-ion universal gate set:}  (Upper panel) Native trapped-ion gates. For a single-ion register, we can apply single-qubit rotations $X(\theta), Y(\theta), Z(\theta)$. For a two-ion register, we can apply entangling MS gates. For an $N$-qubit register of co-trapped ions, the rotations  $X(\theta), Y(\theta)$ act globally on all $N$ qubits, whereas the  $Z(\theta)$ rotations can be addressed individually to the desired qubit subset. Finally, the entangling MS gates also act globally on all $N$ qubits, creating multi-partite entangled states. (Middle panel) The global operations (e.g.~an $X(\theta)$ rotation) can be applied to single qubits of a larger subset by applying spin-echo-type refocusing techniques. (Lower panel) The native trapped-ion gates form a universal gate set. This follows, for instance, from the equivalence of the Pauli and universal gate set (i.e. $S$ gate, Hadamard $H$,  and $CNOT$, generate the  Pauli group, whereas including the $T$ gate instead of  $S=T^2$, leads to a universal gate set~\cite{nielsen-book}) with certain sequences of trapped-ion gates.}
\end{centering}
\end{figure}

Besides being the smallest fully-functional topological qubit, the 7-qubit color code~\cite{bombin-prl-97-180501} permits a transversal bit-wise realization of the entire group of Clifford gate operations (see Fig.~\ref{Fig:gate_set}), which contrasts  the case of the smallest-possible $9$-qubit surface code~\cite{surface_code_9,surface_code_low_d_noise}.  In both cases, a universal set of logical gate operations can be achieved by complementing the Clifford operations with a single non-Clifford gate such as the $T$ gate (see Fig.~\ref{Fig:gate_set}) by magic-state injection \cite{bravyi-pra-71-022316}. Therefore, from a resource-optimization philosophy, we shall model   near-term trapped-ion experiments based on the 7-qubit color code, which can then be extended to larger registers, hosting logical qubits of larger logical distance and increased error robustness, as illustrated in Fig.~\ref{Fig:7qubitCode}.

\subsubsection{The trapped-ion QEC toolbox}
We now describe briefly the main ingredients of the QCCD {\it trapped-ion toolbox} for QEC explored in this work (see Appendix~\ref{sec:appendix_A} for  more details). Building on~\cite{eQual_1qubit,latt_surgery}, we will focus on the  elementary  operations that can be implemented with  high-optical-access (HOA) segmented ion traps in a cryogenic environment~\cite{HOA2_trap} (see Fig.~\ref{Fig:trap_gate_set}). We consider a two-species ion register, such that the elementary quantum information units  can be stored in the electronic states of  one of the species (e.g.~$^{40}$Ca$^+$ optical qubits are encoded in the   ground-state and meta-stable  electronic levels $\ket{0}=\ket{4S_{1/2},-1/2}$ $\ket{1}=\ket{3D_{5/2},-1/2}$, which are  labelled by the principal quantum number and various  orbital/spin angular momenta $\ket{nL_J^{2S+1},M_J}$).  On the other hand,  the remaining species (e.g.~$^{88}$Sr$^+$) shall be exploited for sympathetic cooling to maintain sufficiently-low temperatures of the ion register at certain stages of the QEC cycles. These ions are confined  above a planar segmented trap divided into  manipulation and storage regions, where various elementary operations can be performed:

\begin{figure}[t]
 \begin{centering}
  \includegraphics[width=1.0\columnwidth]{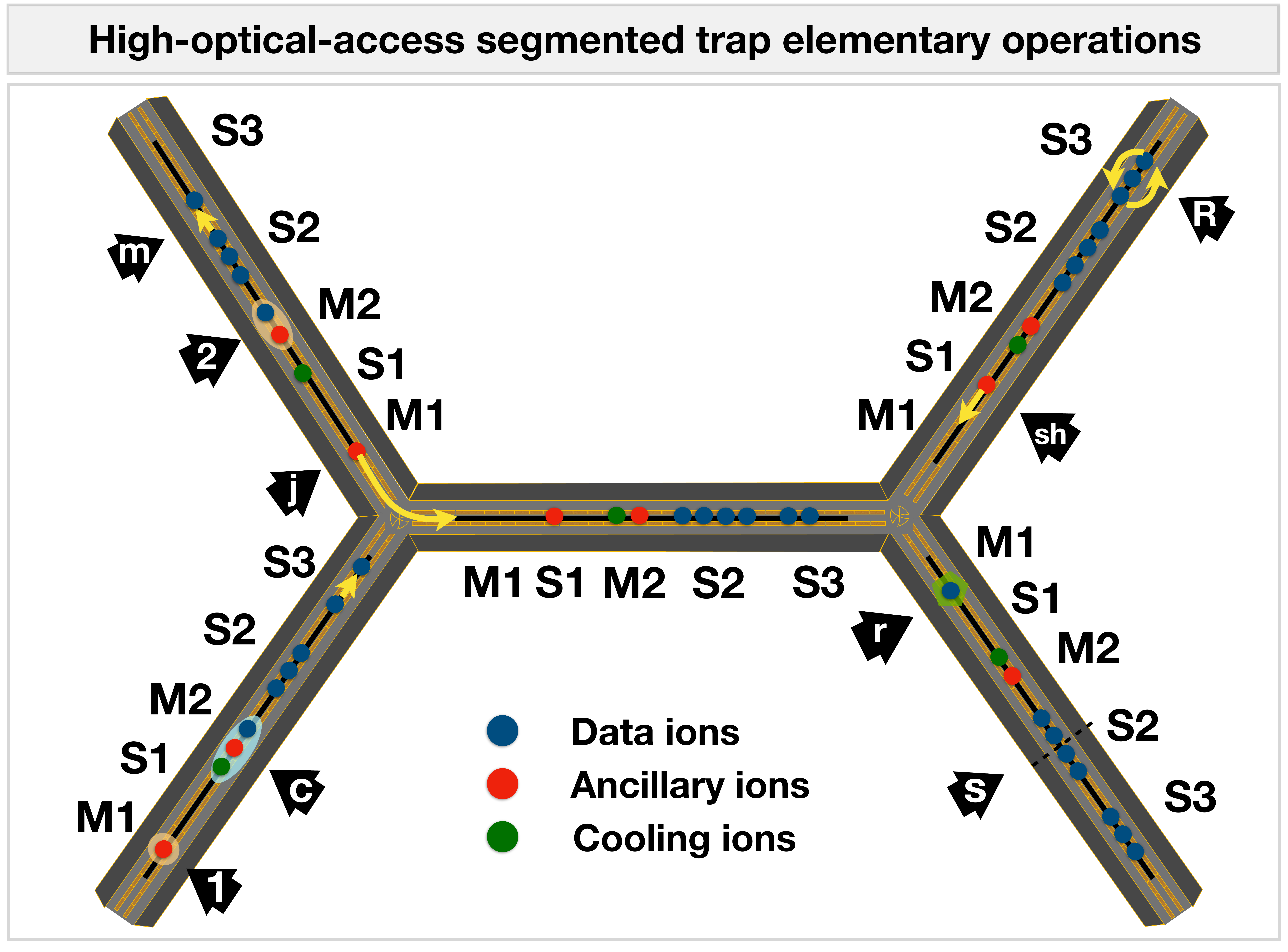}\\
  \caption{\label{Fig:trap_gate_set} {\bf Scheme of the segmented trap and elementary operations:}  We consider a planar trap composed of several arms connected via $\mathsf{Y}$ junctions. These arms  consist of a  linear section divided into  three storage (S1,S2,S3), and two manipulation (M1,M2) zones, each of which contains small  segmented electrodes that allow for ion trapping and control of the motional degrees of freedom.  From left to right, and upper to lower arms, the possible operations considered in this work are (m) merge  sets of ions into a single crystal,  (2) two-qubit entangling gates,(j) shuttling of ion(s) across a junction, (sh) shuttling an ion(s) among neighboring trapping zones, and (R) rotation of an ion crystal. In addition, one can apply (1) single qubit gates, (c) sympathetic laser cool the ion crystal,(r) repump a leaked qubit, and (s) split an ion crystal. The resulting  operations are indicated  by black arrows with the above letters as labels.}
\end{centering}
\end{figure}

{\it (i) Electronic-state manipulation techniques.--} The quantum information can be processed by exploiting various forms of the laser-ion interaction in a given manipulation zone, which yields a universal gate set~\cite{schindler-njp-15-123012}  (see Table~\ref{tab:summary_toolbox}). This gate set contains  $\mathsf{(o1)}$-$\mathsf{(o2)}$ {\it multi-ion entangling gates}  gates  based on the so-called M$\o$lmer-S$\o$rensen (MS) scheme~\cite{molmer-prl-82-1835, PhysRevA.62.022311}, which  allows to generate entanglement between co-trapped ions  mediated by the quantum vibrations of the ion crystal (i.e.~phonons). We note that trapped-ion architectures offer the most accurate entangling gates reported to date~\cite{MS_gates_high_fidelities}, and that current efforts are also being directed towards increasing the gate speed~\cite{MS_gates_fast}.  The corresponding unitary of an $N$-ion entangling gate is parametrised as
\beq
\label{eq:MS_gate}
U_{\rm MS,\phi}(\theta)= \ee^{-\ii \frac{\theta}{4} S_{\phi} ^2},  \hspace{1ex}S_{\phi} = \sum_{i=1} ^N   \cos(\phi) X_i + \sin(\phi) Y_i,
\eeq 
where $X_i=\sigma_i^x$, $Y_i=\sigma_i^y$ and $Z_i=\sigma_i^z$ are expressed in terms of Pauli matrices, and 
$\phi,\theta$ are fully-tunable laser parameters  described in Appendix~\ref{sec:appendix_A}, where a realistic description of the MS scheme in the presence of noise and errors is also presented.

 In the main part of the text, however, we will advocate for a hardware-agnostic language that tries to make  trapped-ion QEC  accessible to non-experts. In particular, the fully-entangling MS gates $U_{\rm MS, 0}(\pi/2)$ ($U_{\rm MS, \pi/2}(\pi/2)$) used in this work, which  generate GHZ-type entangled states such as
\beq
\ket{0_1,0_2,\cdots, 0_N}\to \frac{1}{\sqrt{2}}\bigg(\ket{0_1,0_1,\cdots, 0_N}-\ii\ket{1_1 ,1_2,\cdots, 1_N}\bigg),
\eeq
 will be represented by an abstract circuit analogous to the usual CNOT gate~\cite{nielsen-book}.  We use a vertical string  joining  encircled X (Y) operations for all the  qubits involved in the  gate (see Fig.~\ref{Fig:gate_set}), which are the ions of a single species  co-trapped in the same manipulation zone (see Fig.~\ref{Fig:trap_gate_set}).

In addition to the MS gates, the universal gate set contains $\mathsf{(o3)}$ {\it one-qubit gates} (see Table~\ref{tab:summary_toolbox}).  These gates can either arise from  global rotations around an axis lying in the equatorial plane of the Bloch sphere, which yields the unitary
\beq
\label{eq:global_rotations}
U_{\rm R,\phi}(\theta)=\ee^{-\ii\frac{\theta}{2}S_\phi},
\eeq
  acting simultaneously on all qubits of a certain manipulation zone; or from local rotations about the $z$-axis 
  \beq
  \label{eq:local_rotations}
  U_{{\rm R}_i,z}(\theta)=\ee^{-\ii\frac{\theta}{2}Z_i},
  \eeq
   which can be addressed to the desired $i$-th qubit. Once again, the corresponding microscopic evolution in presence of the main sources of errors are discussed in Appendix~\ref{sec:appendix_A}. These gates will be represented by white rectangular boxes with labels $ X(\theta),Y(\theta)$ for the global rotations, and  $Z(\theta)$ for the local ones (see Fig.~\ref{Fig:gate_set}). As shown in this figure, this trapped-ion native gate set can be used to obtain the more-standard universal gate set of the circuit-based approach to quantum computation~\cite{nielsen-book}, up to  irrelevant global  phases. In a hardware-agnostic spirit, one can use these  relations  to generate any particular unitary operation on an $N$-qubit trapped-ion register.
   
   In addition to these operations,  to translate any quantum-information protocol into the trapped-ion hardware, we note that the ions have a so-called closed cycling transition~\cite{schindler-njp-15-123012} that leads to $\mathsf{(o4)}$ {\it  projective measurements} in the z-basis by collecting a state-dependent fluorescence due to the emitted photons from the cycling transition, and $\mathsf{(o5)}$ {\it qubit initialization/reset}  by optical pumping.

\begin{table}
  \centering
  \begin{tabular}{|l |c|c|c|c|} \hline\hline
    Operation & Current & Current & Anticipated & Anticipated  \\
      &  duration & infidelity  &  duration &   infidelity \\\hline
        $\mathsf{(o1)}$  Two-qubit    & 40$\mu$s & $1 \cdot 10^{-2}$ & 15$\mu$s & $2 \cdot 10^{-4}$ \\ 
       MS gate  &  & &  &  \\\hline
      $\mathsf{(o2)}$ Five-qubit   & 60$\mu$s & $5 \cdot 10^{-2}$ & 15$\mu$s & $1 \cdot 10^{-3}$ \\ 
      MS gate  &  & &  &  \\\hline
      $\mathsf{(o3)}$ One-qubit gate & 5$\mu$s & $5 \cdot 10^{-5}$ & 1$\mu$s & $1 \cdot 10^{-5}$ \\ \hline
     $\mathsf{(o4)}$ Measurement & 400$\mu$s & $1 \cdot 10^{-3}$ & 30$\mu$s & $1 \cdot 10^{-4}$ \\ \hline
      $\mathsf{(o5)}$ Qubit reset & 50$\mu$s & $5 \cdot 10^{-3}$  & 10$\mu$s &$5 \cdot 10^{-3}$ \\ \hline
      $\mathsf{(o6)}$ Re-cooling & 400$\mu$s & $\bar{n} < 0.1$ & 100$\mu$s & $\bar{n} < 0.1$ \\ \hline
      $\mathsf{(o7)}$ Ion shuttling  & 5$\mu$s & $\bar{n} < 0.1$ & 5$\mu$s & $\bar{n} < 0.1$ \\ \hline
       $\mathsf{(o8)}$ Ion  split/merge  & 80$\mu$s & $\bar{n} < 6$ & 30$\mu$s & $\bar{n} < 1$ \\ \hline
        $\mathsf{(o9)}$ Ion  rotation  & 42$\mu$s & $\bar{n} < 0.3$ & 20$\mu$s & $\bar{n} < 0.2$ \\ \hline
          $\mathsf{(o10)}$ Junction  & 100$\mu$s & $\bar{n} < 3$ & 200$\mu$s & $-$ \\
                 crossing (per ion)  &  & &  &  \\\hline
                 $\mathsf{(o11)}$ Leakage & 60$\mu$s & $5 \cdot 10^{-3} $ & 20$\mu$s & $5 \cdot 10^{-3}$ \\
                 repumping   &  & &  &  \\\hline
  \end{tabular}
  \caption{{\bf Extended trapped-ion QEC toolbox}. Description of current and future trapped-ion capabilities for a QCCD approach to FT QEC, following~\cite{eQual_1qubit}. We include the duration and infidelity of operations acting on the internal degrees of freedom, and the duration and final mean number of phonons in the longitudinal center-of-mass mode for the operations involving the external degrees of freedom of the ions.}
  \label{tab:summary_toolbox}
\end{table}

{\it (ii) Ion-crystal-reconfiguration techniques.--} The trapped-ion QEC toolbox  includes techniques to control the external and motional degrees of freedom of the ion crystal. In particular, we shall exploit the other species, e.g.~$^{88}$Sr$^+$ ions, for $\mathsf{(o6)}$ {\it re-cooling of the ion crystal} using sympathetic  laser cooling techniques. In this way, one can cool the vibrational mode that is used as a quantum data bus to generate entanglement, prior to any entangling gate (see Table~\ref{tab:summary_toolbox}), such that  high fidelities can be still be achieved after the ion crystal has gone through a sequence of reconfiguration operations. 

These crystal reconfiguration operations, which heat the vibrational modes, can be applied in both manipulation and storage regions, as they  require control over the  trapping potentials  but no lasers are involved. We consider the following elementary operations: $\mathsf{(o7)}$ {\it fast  shuttling of ions  or small crystals} across different segments of a single arm of the trap; $\mathsf{(o8)}$ {\it fast splitting and merging of ion crystals}; and $\mathsf{(o9)}$ {\it fast  swapping of pairs of ions and rotations of small  crystals} around a reflexion axis. Although these operations do not appear explicitly in the abstract circuits of Fig.~\ref{Fig:gate_set}, they are a fundamental ingredient in the microscopic schedules that need to be realized for the implementation of a particular QEC protocol. 

Following the spirit of this work, a hardware-agnostic language  would not require to know the particular pulse sequences required to perform these microscopic operations (see Fig.~\ref{Fig:trap_gate_set}), nor the schedules of such operations that must be applied to perform a sequence of gates for particular stages of the QEC protocol. Accordingly, some of these details   are  relegated to  Appendix~\ref{sec:appendix_B}.  For non-experts in trapped-ion physics, we only need to know that these reconfiguration operations do not act on the  quantum information encoded in the electronic states. Therefore, we only have to incorporate the effect of  the environmental noise that affects the  qubits during the time that these operations take (see Table~\ref{tab:summary_toolbox}). We will describe the corresponding   error models in the following subsection.

\begin{figure}[t]
 \begin{centering}
  \includegraphics[width=1.0\columnwidth]{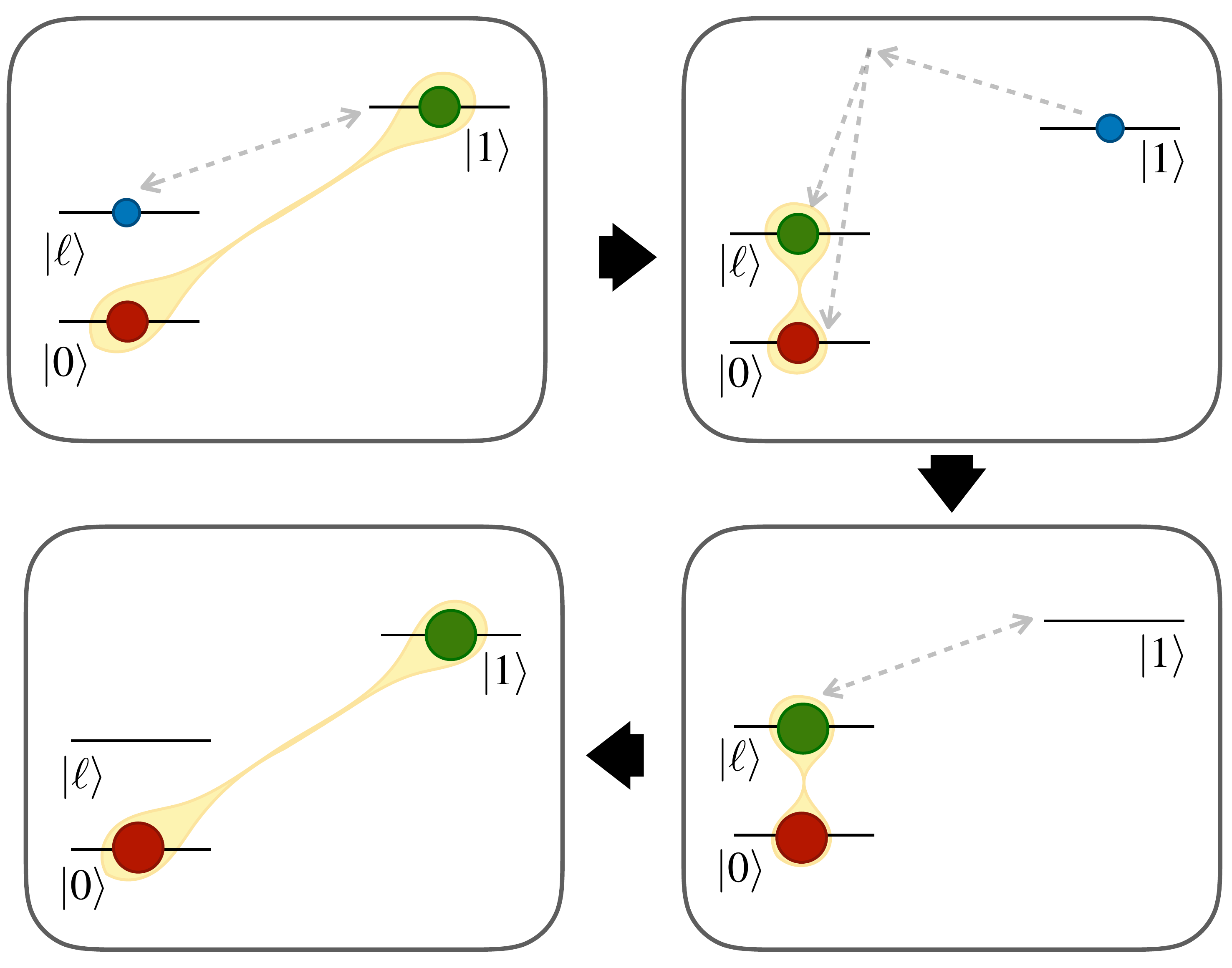}\\
  \caption{\label{Fig:repumping_scheme} {\bf Repumping scheme for leakage errors:} Due to spontaneous decay,  the population of the leaked level can grow, but initially-vanishing coherences with the qubit levels will not build up. In the first step (upper-left panel), a $\pi$-pulse brings the population of the leaked state $\ket{\ell}$ to the metastable level, coherently hiding the qubit state in the ground-state manifold. In a second step (upper-right panel), by driving a dipole-allowed transition between the metastable level and higher-excited levels, the leaked population can be pumped back to the computational subspace via spontaneous decay onto the ground-state manifold. Note that the coherences initially present between the populations stored in the ground-state manifold are not affected in this process. In a third step (lower-right panel), a $\pi$-pulse brings the hidden and repumped qubit state back to the computational subspace (lower-left panel).}
\end{centering}
\end{figure}

So far,  $\mathsf{(o1)}$-$\mathsf{(o9)}$ form the   QEC toolbox  used to assess the progress of near-term trapped-ion QEC~\cite{eQual_1qubit}. As discussed in~\cite{latt_surgery}, the extensibility towards larger registers  requires  including $\mathsf{(o10)}$ {\it transport of ions across junctions}~\cite{x_junction_transport,Y_junction_transport} connecting different arms of the  trap (see Fig.~\ref{Fig:gate_set}). In this work, we include a new and  important ingredient in this toolbox. As discussed below, a realistic modeling of the microscopic trapped-ion noise requires  considering  leakage of the electronic state out of the computational subspace. Due to the choice of  $^{40}$Ca$^+$ optical qubits, the metastable qubit state $\ket{1}=\ket{3D_{5/2},-1/2}$ can spontaneously decay into the ground-state qubit state $\ket{0}=\ket{4S_{1/2},-1/2}$, resulting in an amplitude damping within the computational subspace. However, there is a also a finite branching ratio for the decay into a different ground-state level  $\ket{\ell}=\ket{4S_{1/2},+1/2}$, giving rise to leakage out of the qubit subspace. As discussed below, although the trapped-ion QEC  protocols~\cite{eQual_1qubit,latt_surgery} can  cope with amplitude-damping errors, the leakage cannot be counteracted by the error correction, such that  the leaked population will accumulate as the protocol proceeds, compromising the usefulness of the QEC. 

To counteract this detrimental effect, we  include a $\mathsf{(o11)}$ {\it repumping pulse sequence} that can bring the population of the electronic state back to the computational subspace without affecting the coherences. The main idea is that, contrary to the bare-leakage  errors, which are uncorrectable, the repumped-leakage  errors behave effectively like amplitude damping within the computational space, and  can thus be corrected by subsequent QEC cycles (see Fig.~\ref{Fig:repumping_scheme}). In a hardware-agnostic language, knowledge of the laser pulse sequence for the repumping detailed in Appendix~\ref{sec:appendix_A} is not required. Instead, we need to describe a microscopic noise model describing how a faulty repumping  alters the amplitudes  of the quantum state, which will be described in the following section.

\subsection{Realistic microscopic error models}
\label{sec:noise_models}

As emphasized in the introduction,   theoretical predictions about the performance of   QEC  can differ substantially depending on the assumptions about the experimental capabilities, and the models used to describe the effects of environmental noise and/or  experimental imperfections. A clear example of this trend is the existence of a wide range of FT thresholds~\cite{dennis-j-mat-phys-43-4452,raussendorf-prl-98-190504,threshold_color_code} depending on the noise model~\cite{threshold_color_code_circuit_noise}. Therefore,  meaningful assessments of the near-term prospects of various technologies to demonstrate the break-even point of beneficial QEC require a realistic microscopic noise modeling. We believe this type of studies are important to guide near-future technological developments that must be accomplished in the progress towards FT quantum computers.

 In~\cite{eQual_1qubit,latt_surgery}, a microscopic description of the possible technical imperfections and environmental sources of noise has been presented. In this section, we build on these efforts to update the microscopic description providing, to the best of our knowledge, the most complete and realistic error model for near-term QEC that can be found in any candidate platforms explored to date. Following our hardware-agnostic effort, we describe the main properties of the noise model for non-experts, and relegate the details to Appendix~\ref{sec:appendix_A}.

\subsubsection{Coherent errors  for single-qubit gates}

Let us consider the global~\eqref{eq:global_rotations} and local~\eqref{eq:local_rotations} single-qubit rotations. As advanced in the introduction, and described in detail  in Appendix~\ref{sec:appendix_A}, the single-qubit gates~\eqref{eq:global_rotations}-\eqref{eq:local_rotations} are defined by parameters $\theta,\phi$ that depend on the laser-beam intensity and phase, which may fluctuate around the target value $\theta\to\theta(t)=\theta+\delta\theta(t)$, $\phi\to\phi(t)=\phi+\delta\phi(t)$. The corresponding global rotations used in Fig.~\ref{Fig:gate_set}, must be substituted by 
\beq
\label{noisy_carrier}
\begin{split}
X(\theta)&\to \hat{X}(\theta)=\ee^{-\ii\frac{\theta(t)}{2}{S}_{\delta\phi(t)}}, \\
 Y(\theta)&\to \hat{Y}(\theta)=\ee^{-\ii\frac{\theta(t)}{2}{S}_{\delta\phi(t)+\pi/2}},
 \end{split}
\eeq  
where  we have assumed that the phase fluctuations occur on a much slower time-scale than the intensity fluctuations.
Accordingly, the gates  suffer from under- or over-rotations due to laser intensity fluctuations leading to  non-zero $\delta\theta(t)$; while  the phase drifts $\delta\phi(t)$ yield fluctuations in the rotation axis
\beq
\begin{split}
{S}_{\delta\phi(t)}&=\sum_i \cos(\delta\phi(t)){X}_i + \sin(\delta\phi(t)){Y}_i,\\
{S}_{\delta\phi(t)+\pi/2}&=\sum_i \cos(\delta\phi(t)){Y}_i - \sin(\delta\phi(t)){X}_i.\\
\end{split}
\eeq
 Finally,  local rotations in Fig.~\ref{Fig:gate_set} should be substituted by 
 \beq
 \label{eq:noisy_local_rotations}
 Z(\theta)\to \hat{Z}(\theta)=\ee^{-\ii\frac{\theta(t)}{2}Z}.
 \eeq
  In this case, since they arise from two-photon ac-Stark shifts, the typically-slow phase drifts have no effect on the rotation axis. Conversely, intensity fluctuations  can again   yield under/over-rotations via $\delta\theta(t)$.

In contrast to typical Kraus-map modelling of noise in QEC, our  faulty gates are  not characterized by a single error rate, but instead by two fluctuating functions $\delta\theta(t),\delta\phi(t)$ described by a  random process with parameters fixed by experimental considerations~\cite{random_processes, ou_random_process}.  A hardware-agnostic approach only requires prior knowledge of these fluctuating functions, such that the faulty gates $\hat{X}(\theta),\hat{Y}(\theta),\hat{Z}(\theta)$ can be simulated  in  highly-parallelized full-wave-function  simulations, as  discussed below, going in this way beyond the  typical  Pauli errors explored in QEC~\cite{simulation_clifford}.

\subsubsection{Error model for two-qubit entangling gates}

 Let us now consider the  MS gates~\eqref{eq:MS_gate} for a two-ion crystal, which are  essential ingredients in the FT stabilizer readout for trapped-ion QEC. The MS scheme creates a state-dependent force by exploiting the laser-ion interaction in the regime of resolved phonon sidebands~\cite{molmer-prl-82-1835,PhysRevA.62.022311}. This force, which displaces the ions along trajectories that depend on their electronic state, can also yield a collective geometric phase responsible for the entangling gate~\cite{review_sd_force}. We note that the underlying laser-driven qubit-phonon dynamics can lead to various sources of errors, such as { motional errors}  (i.e.~residual spin-phonon entanglement of spectator modes and Debye-Waller fluctuations of the Rabi frequencies)~\cite{PhysRevA.62.022311},  {  dephasing} (i.e.~decoherence due to fluctuations of global magnetic fields), and fluctuations of the laser intensity/phase.  The time-evolution of the two qubits subjected to the laser-ion interaction is
\beq
\label{eq:MS_gate_ev}
\rho(t_{\rm g})\!={\rm Tr}_{\rm ph}\!\left(
U_{\rm g}\rho_0U^\dagger_{\rm g}
\right)\!,\hspace{1ex} U_{\rm g}\!=\ee^{-\ii t_{\rm g}H_0}\mathsf{T}\left\{\ee^{-\ii\int_0^{t_{\rm g}}{\rm d}t' H_{\rm int}(t')}\!\right\},
\eeq
where ${\rm Tr}_{\rm ph}$ represents the partial trace over the phonons, $t_{\rm g}$ is the gate time,  $H_0$ contains the independent dynamics of the electronic and vibrational degrees of freedom, $\mathsf{T}$ is the time-ordering operator,, and $H_{\rm int}(t')$ describes the laser-ion interactions in the resolved-sideband regime~\cite{roos_gates}, including the above potential sources of errors (see Appendix~\ref{sec:appendix_A}). In this microscopic description, $\rho_0$ contains the initial qubit state and a thermal state for the vibrations with typical phonon numbers described in Table~\ref{tab:summary_toolbox}, whereas the final qubit state is obtained by tracing over the vibrational states on the time-evolved state. 

We  solve numerically the qubit-phonon dynamics~\eqref{eq:MS_gate_ev}, and  perform process tomography~\cite{process_tomography} to express the final qubit state as the result of a   generic quantum channel
\beq
\label{eq:MS_channel}
\rho(t_{\rm g})=\sum_np_{n}K_n^{\phantom{\dagger}}\rho_0K_n^{\dagger}, \hspace{3ex}\sum_np_{n}K_n^\dagger K_n^{\phantom{\dagger}}=\mathbb{I}.
\eeq
Here,  $K_n$ are   two-qubit Kraus operators, and
$p_{n}$ are their corresponding probabilities, already averaged over the stochastic processes that describe dephasing and laser-parameter fluctuations. Note that all the different sources of error introduced above will result in a set $\{p_{n}\}_{n=1}^{16}$, where $p_{1}\approx1$ corresponds to a Kraus operator $K_1$ close to the ideal MS gate~\eqref{eq:MS_gate}, whereas the remaining weights $\{p_{2},p_{3},\cdots\}$ correspond to the most-likely errors and we use a  decreasing ordering (see Appendix~\ref{sec:appendix_B}). It turns out that these weights decay very fast, and that the more-relevant errors occur as single-qubit Pauli operators in the same basis as the MS gate.

For the hardware-agnostic description, once the set $\{{p}_n,{K}_n, \forall n: {p}_n>p_{\rm trunc}\}$ is given, one  can  readily incorporate it  in a Monte Carlo approach full-wave-function simulation.  For a pure-state Monte Carlo evolution, we need to generate random numbers $ r \in[0,1]$ and apply the numerically generated ${K}_n$ if $r$ falls in the respective probability interval, $\sum_{k}^{n-1} {p}_{k} \leq r < \sum_{k}^{n} {p}_{k}$, where $p_0=0$. In this way, one  samples over all the relevant Kraus operators, such that the stochastic average yields the noisy MS gate.

\subsubsection{Amplitude damping and qubit leakage}
\label{sec:leakage}

Typically, environmental dephasing is considered to be the main source of  noise affecting idle trapped-ion qubits~\cite{eQual_1qubit,latt_surgery}. However, near-term technical improvements are expected to reach  coherence times close to the limit of $T_2=2T_1\approx2.2\,$s, such that  amplitude damping from  the metastable state $\ket{1}=\ket{3D_{5/2},-1/2}$  into  $\ket{0}=\ket{4S_{1/2},-1/2}$ must be also considered. Moreover, the spontaneous decay can also populate a ground-state level  that does not belong to the computational subspace $\ket{\ell}=\ket{4S_{1/2},+1/2}$, leading to the aforementioned  leakage errors. Given the Markovian nature of the electromagnetic environment responsible for this spontaneous decay, and the typical separations between co-trapped ions forming a crystal, we can directly rule out  effects from temporal~\cite{atom_photon_int} and spatial~\cite{superradiance_ions} correlations in the spontaneous decay.

\begin{figure}
\centering
\includegraphics[width=1\columnwidth]{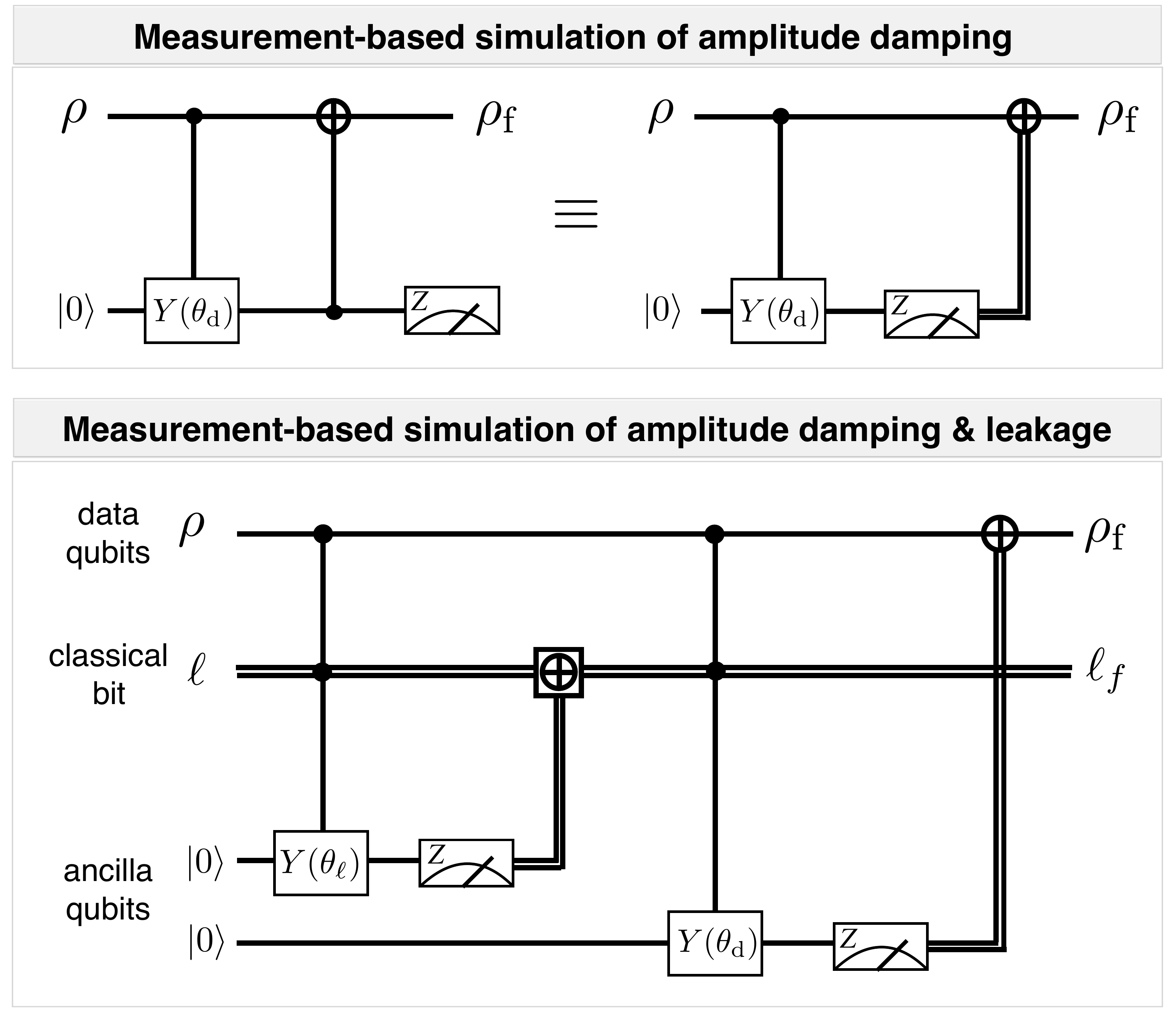}
\caption{ {\bf  Circuit-based simulation of amplitude damping and leakage:}  (Upper panel) The amplitude damping channel can be obtained from a circuit that couples the data qubit to an ancillary qubit using a pair of entangling gates, a conditional $Y$-rotation and a CNOT, followed by a measurement of the ancillary qubit. This is fully equivalent to the circuit where the second entangling gate is substituted by a single-qubit rotation that is applied conditional on the result of the ancilla measurement, as depicted by a double line where the classical information determines the subsequent single-qubit rotation. {(Lower panel)} Circuit to simulate the amplitude damping and leakage, including an auxiliary classical bit to account for leakage events, and a pair of ancillary qubits. These qubits are entangled to the data qubit via controlled $Y$-rotations that depend on the state of both the classical and quantum bits, such that one can simulate simultaneously the damping/leakage via measurement-dependent operations.}
\label{damping_leakage_circuit}
\end{figure}

Let us note that the amplitude-damping channel $\rho(t)=\sum_nL_n\rho L_n^\dagger$ with  $L_0=\ket{0}\bra{0}+\sqrt{1-p_{\rm ad}(t)}\ket{1}\bra{1}$ and $L_1=\sqrt{p_{\rm ad}(t)}\ket{0}\bra{1}$, can be incorporated in a circuit-based simulation by means of an auxiliary qubit~\cite{nielsen-book}. This ancilla qubit must be entangled with the data qubit via a controlled rotation of angle $\theta_{\rm d}$,  a subsequent CNOT gate, and finally measured in the computational basis (see upper panel  of  Fig.~\ref{damping_leakage_circuit}). In this simplified case, the  rotation angle is fixed by the spontaneously-decayed population $\theta_{\rm d}=2{\rm arcsin}\left((1-{\rm exp}(-t_{\rm id}/T_1))^{1/2}\right)$, such that the amplitude-damping error rate is $p_{\rm ad}(t)=\sin^2(\theta_{\rm d}/2)$.

To incorporate the possible leakage  $\rho(t)=\sum_nL_n\rho L_n^\dagger$ with  $L_0=\ket{0}\bra{0}+\ket{\ell}\bra{\ell}+\sqrt{1-p_{\rm ad}(t)-p_{\ell}(t)}\ket{1}\bra{1}$, $L_1=\sqrt{p_{\rm ad}(t)}\ket{0}\bra{1}$, and $L_2=\sqrt{p_{\ell}(t)}\ket{\ell}\bra{1}$, we must generalize the circuit to describe this process.
 We  consider that the initial state has no coherences between the computational states and the leaked level. Since the leaked population can only increase  by incoherent spontaneous decay, such coherences cannot be build up,  and we can therefore use at all times one auxiliary classical bit per ion to store the information about the leaked level, i.e.~$\ell=0$ ($\ell=1$) if the qubit has (has not) leaked. As depicted in the lower panel of  Fig.~\ref{damping_leakage_circuit}, the circuit-based simulation of  simultaneous amplitude damping and leakage requires a couple of ancillary qubits, and a pair of controlled rotations that are also conditioned on the qubit not having leaked into $\ket{\ell}$. As described in detail in Appendix~\ref{sec:appendix_A}(c), the rotation angles must now be set to
\beq
\label{eq:decay_angles}
\begin{split}
\theta_{\ell}&=2\arcsin\sqrt{\frac{\Gamma' \left(1-\ee^{-(\Gamma+\Gamma')t}\right)}{\Gamma+\Gamma'}},\\
\theta_{\rm d}&=2\arcsin\sqrt{\frac{\Gamma\left(1-\ee^{-(\Gamma+\Gamma')t}\right)}{\Gamma+\Gamma'\ee^{-(\Gamma+\Gamma')t}}},
\end{split}
\eeq
where we have introduced the branching ratio of spontaneous decay into the leaked level $\Gamma'/\Gamma$, which is typically small (i.e. $\Gamma'/\Gamma\approx4/9$ for our particular  $^{40}$Ca$^+$optical qubits~\cite{roos_thesis}). We note that in the numerical simulation it is sufficient to use in total merely two ancillary qubits, independently of the number $N$ of ions, to simulate leakage dynamics on the entire register of trapped-ion qubits. 

Following our hardware-agnostic goal, one can model the spontaneous decay by simply implementing the circuit of Fig.~\ref{damping_leakage_circuit} with the angles given by Eq.~\eqref{eq:decay_angles}. Let us now discuss how the subsequent operations are affected by the leakage. For single qubit gates, if the ion is indeed in the $\ket{\ell}$ level, the lasers will only cause an off-resonant ac-Stark shift  that can be fully neglected as there are no coherences between the leaked and un-leaked states. For the two-qubit MS gates, the situation is a bit more involved. In case both qubits  have  leaked,  the lasers responsible for the gate will be highly off-resonant leading to   irrelevant ac-Stark shifts that can be neglected as before. On the other hand, if only one of the qubits has leaked, the lasers will still be near-resonant with the sidebands  of the un-leaked qubit. This qubit will evolve under a state-dependent force, as discussed in a previous subsection,  realizing a trajectory in phase space, during which the spin is entangled with the motion. Note, however, that the timing of the gate still guarantees that the  phase-space trajectory will be closed (i.e. it is always an integer multiple of the detuning, regardless of one of the ions not participating in the MS gate). Hence, to leading order in our MS-gate error model, the un-leaked qubit simply develops a closed-trajectory that is equivalent to the identity operator. For the subsequent elementary operations after the idle periods, if the classical bit signals leakage $\ell=0$,  single-qubit gates simply act as the identity, while two-ion MS gates act as the identity on the qubits involved in the gate. Therefore, in a hardware-agnostic language, the elementary operations involving a leaked level  correspond to the identity.
 
In addition to these improved error models, we also use dephasing noise in the Markovian regime, as well as a bit-flip channel to model imperfect qubit initialization and readout~\cite{eQual_1qubit,latt_surgery}, forming altogether our microscopic trapped-ion error model.

\subsection{Flag- versus cat-based  stabilizer readout}

We now describe two possible strategies for one of the crucial operations in active QEC: the readout of the plaquette stabilizers~\eqref{eq:stabilizers}. Obtaining $-1$ measurement values signal  the occurrence of  errors, which  take the state out of the code subspace $\ket{\Psi}\notin\mathcal{V}_{\rm code}$. The role of the active QEC strategies is to devise a strategy to: {\it (a) } perform these measurements without affecting the quantum information encoded in the system, {\it (b)} avoid the uncontrolled propagation of errors using  FT constructions of the corresponding circuits, and {\it (c)} devise decoders that allow to infer the most-likely error for a given set of stabilizer measurements. Until very recently~\cite{flag_based_readout},   there were three main  strategies for  FT stabilizer readout~\cite{shor_ft_qec,steane_ft,knill_ft}. 

Regarding the trapped-ion  experimental capabilities, the required  resources  can be minimized   with a   Shor-type  readout~\cite{eQual_1qubit}, whereby the non-demolition measurement makes use of ancillary qubits  prepared in  entangled  cat states  to avoid the proliferation of errors during the FT readout~\cite{shor_ft_qec,aliferis_ft_qec}.  However, the preparation and certification of highly-entangled cat states is still a resource-intensive requirement. As discussed in~\cite{latt_surgery}, the resources can be optimized further by moving into a flag-based readout scheme~\cite{flag_based_readout,flag_readout_arb_distance}, whereby the  cat states  are substituted   by  a so-called  {flag  qubit}, which is operated  in combination with a bare syndrome  qubit onto which the stabilizer information gets mapped. One of the goals of the present work is to perform a comparative numerical study of cat- and flag-based approaches using the realistic trapped-ion error model. We thus start by describing these two different FT strategies.
 
\begin{figure}[t]
 \begin{centering}
  \includegraphics[width=1\columnwidth]{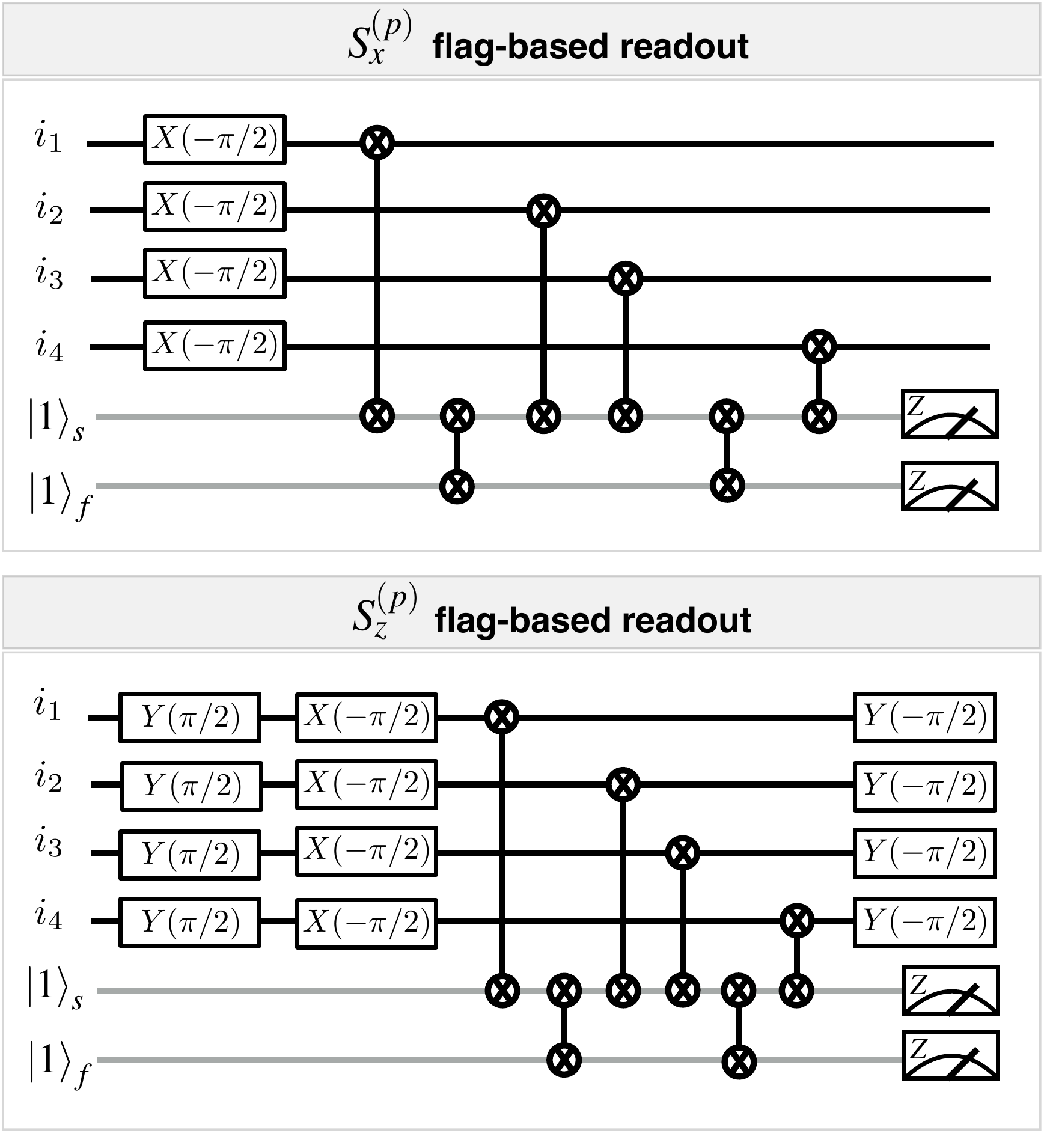}\\
  \caption{\label{Fig:flag_readout_ions} {\bf  Flag-based stabilizer readout with the trapped-ion universal gate set:}  Trapped-ion circuit for the  flag-based measurements of $S_x^{(p)},S_z^{(p)}$. The upper qubits labelled as $i_1,i_2,i_3,i_4$ correspond to one of the  stabilizers~\eqref{eq:stabilizers} of a particular plaquette (see Fig.~\ref{Fig:7qubitCode}), and are represented by black  lines. The lower qubits $s,f$, represented by grey lines, correspond to the syndrome and flag ancillary qubits, respectively, which are initialized in the $\ket{1}$ state. Note that the $z$-type stabiliser readout is obtained from the $x$-stabiliser readout by a simple $y$-rotation, and there is no need to change control and target as occurs for CNOT-based schemes. }
\end{centering}
\end{figure}

 In the flag-based approach, the flag qubit is coupled to the syndrome qubit by a pair of MS entangling gates (see Fig.~\ref{Fig:flag_readout_ions}), which serves to detect the  cascading of correlated errors into the data qubits. The combination of the flag readout with subsequent stabilizer measurements allows to  identify and correct the most-likely  error. Let us recall that errors propagate through the entangling MS gates acting on the $i,j$ pair of qubits as follows: $U_{\rm MS,0}(\pi/2)X_i=X_iU_{\rm MS,0}(\pi/2)$, $U_{\rm MS,0}(\pi/2)X_j=X_jU_{\rm MS,0}(\pi/2)$, $U_{\rm MS,0}(\pi/2)Z_i=Y_iX_jU_{\rm MS,0}(\pi/2)$, and  $U_{\rm MS,0}(\pi/2)Z_j=X_iZ_jU_{\rm MS,0}(\pi/2)$~\cite{eQual_1qubit}. Using the rules for   the  propagation, together with the straightforward  rotations by   single-qubit gates, one can ascertain that an error has indeed occurred  whenever the flag is triggered (i.e. projective measurement in the $z$-basis $M_f=-1$). By performing a subsequent measurement of the three conjugate stabilizers, one can  determine and correct  the most-likely error including the potentially-dangerous correlated errors (see Table~\ref{table_decoding_flags}). Note that these subsequent measurements can be realized using the un-flagged versions of the circuits (i.e.~using a bare syndrome qubit) while maintaining the fault tolerance at level-1 (i.e. the correcting power of the 7-qubit code, namely $t=1$, is not compromised by the syndrome extraction circuits, which maintain fault tolerance at this level and do not allow errors to cascade). The reason is that since the flag has already been triggered, and the 7-qubit color code   can only cope with  a single error, the   only correctable events  are those where  subsequent gates do not introduce  additional errors, such that fault-tolerance can be attained using bare ancillas. If, on the other hand, the flag is not triggered but the stabiliser signals an error $-1$, we know that an error must have occurred on a single qubit  at FT level-1, such that we can again measure the remaining stabilizers with un-flagged circuits to find which single-qubit error is the most likely one.

\begin{table*}
\begin{center}
\begin{tabular}{|c|c|c|c|c|c|c|c|}
\hline
\multicolumn{8}{|c|}{Chao-Reichardt flag-based readout of $X$-type stabilizers}\\
\hline
\multicolumn{2}{|c|}{No flag triggered $M_f=+1$} &\multicolumn{2}{c|}{  $S_x^{(1)}$ flag triggered $M_f=-1$}& \multicolumn{2}{c|}{ $S_x^{(2)}$ flag triggered $M_f=-1$}&\multicolumn{2}{c|}{ $S_x^{(3)}$ flag triggered $M_f=-1$}\\
\hline
Syndrome & Error & Syndrome  & Error & Syndrome & Error & Syndrome  & Error \\
 $\left(S_x^{(1)},S_x^{(2)},S_x^{(3)}\right)$ &  correction & $\left(S_z^{(1)},S_z^{(2)},S_z^{(3)}\right)$  &  correction & $\left(S_z^{(1)},S_z^{(2)},S_z^{(3)}\right)$ & correction  & $\left(S_z^{(1)},S_z^{(2)},S_z^{(3)}\right)$ & correction \\
\hline
 \hspace{2ex}$(+1,+1,+1)$\hspace{2ex} & $\mathbb{I}$ & $\hspace{2ex}(+1,+1,+1)\hspace{2ex}$  & $X_f$  & $(+1,+1,+1)$ & $X_f$ & $(+1,+1,+1)$ & $X_f$ \\
\hline
$(+1,+1,-1)$ &\hspace{2ex}  $Z_7$\hspace{2ex} &  $(+1,+1,-1)$& $X_7X_f$&$\hspace{2ex}(+1,+1,-1)\hspace{2ex}$ &\hspace{2ex}  $X_5X_6$  \hspace{2ex} &$\hspace{2ex} (+1,+1,-1) \hspace{2ex}$ & $X_7$  \\
\hline
$(+1,-1,+1)$ &   $Z_5$ &$(+1,-1,+1)$& $X_3X_4$ & $(+1,-1,+1)$ &   $X_5X_f$    &$(+1,-1,+1)$ &\hspace{2ex}  $X_6X_7$ \hspace{2ex} \\
\hline
$(+1,-1,-1)$ &  $Z_6$ &  $(+1,-1,-1)$ &  $X_6X_f$  &$(+1,-1,-1)$ & $X_6$ & $(+1,-1,-1)$ &  $X_6X_f$    \\
\hline
$(-1,+1,+1)$ & $Z_1$ &$(-1,+1,+1)$  & $X_1$ & $(-1,+1,+1)$ &  $X_1X_f$ & $(-1,+1,+1)$  &  $X_1X_{f}$    \\
\hline
$(-1,+1,-1)$ &  $Z_4$ &  $(-1,+1,-1)$ & $X_4$  & $(-1,+1,-1)$ &  $X_4X_f$   &  $(-1,+1,-1)$ &  $X_4X_f$    \\
\hline
$(-1,-1,+1)$ & $Z_2$ &   $(-1,-1,+1)$ &  $X_2X_f$  &$(-1,-1,+1)$ & $X_2$ & $(-1,-1,+1)$ &  $X_2X_f$   \\
\hline
$(-1,-1,-1)$ & $Z_3$ &   $(-1,-1,-1)$ &  $X_3X_f$   & $(-1,-1,-1)$ &  $X_3X_f$    &$(-1,-1,-1)$ & $X_3$  \\
\hline
\multicolumn{8}{c}{}\\
\hline
\multicolumn{8}{|c|}{Chao-Reichardt flag-based readout of $Z$-type stabilizers}\\
\hline
\multicolumn{2}{|c|}{No flag triggered $M_f=+1$} &\multicolumn{2}{c|}{  $S_z^{(1)}$ flag triggered $M_f=-1$}& \multicolumn{2}{c|}{ $S_z^{(2)}$ flag triggered $M_f=-1$}&\multicolumn{2}{c|}{ $S_z^{(3)}$  flag triggered $M_f=-1$}\\
\hline
Syndrome & Error & Syndrome  & Error & Syndrome & Error & Syndrome  & Error \\
 $\left(S_z^{(1)},S_z^{(2)},S_z^{(3)}\right)$ & correction & $\left(S_x^{(1)},S_x^{(2)},S_x^{(3)}\right)$  & correction  & $\left(S_x^{(1)},S_x^{(2)},S_x^{(3)}\right)$ &  correction & $\left(S_x^{(1)},S_x^{(2)},S_x^{(3)}\right)$ & correction \\
\hline
 \hspace{2ex}$(+1,+1,+1)$\hspace{2ex} &  $\mathbb{I}$ & $\hspace{2ex}(+1,+1,+1)\hspace{2ex}$  & $Z_f$  & $(+1,+1,+1)$ & $Z_f$ & $(+1,+1,+1)$ & $Z_f$ \\
\hline
$(+1,+1,-1)$ &\hspace{2ex}  $X_7$\hspace{2ex} &  $(+1,+1,-1)$& $Z_7Z_f$&$\hspace{2ex}(+1,+1,-1)\hspace{2ex}$ &\hspace{2ex}  $Z_5Z_6$  \hspace{2ex} &$\hspace{2ex} (+1,+1,-1) \hspace{2ex}$ & $Z_7$  \\
\hline
$(+1,-1,+1)$ &   $X_5$ &$(+1,-1,+1)$& $Z_3Z_4$ & $(+1,-1,+1)$ &   $Z_5Z_f$    &$(+1,-1,+1)$ &\hspace{2ex}  $Z_6Z_7$ \hspace{2ex} \\
\hline
$(+1,-1,-1)$ &  $X_6$ &  $(+1,-1,-1)$ &  $Z_6Z_f$  &$(+1,-1,-1)$ & $Z_6$ & $(+1,-1,-1)$ &  $Z_6Z_f$    \\
\hline
$(-1,+1,+1)$ & $X_1$ &$(-1,+1,+1)$  & $Z_1$ & $(-1,+1,+1)$ &  $Z_1Z_f$ & $(-1,+1,+1)$  &  $Z_1Z_{f}$    \\
\hline
$(-1,+1,-1)$ &  $X_4$ &  $(-1,+1,-1)$ & $Z_4$  & $(-1,+1,-1)$ &  $Z_4Z_f$   &  $(-1,+1,-1)$ &  $Z_4Z_f$    \\
\hline
$(-1,-1,+1)$ & $X_2$ &   $(-1,-1,+1)$ &  $Z_2Z_f$  &$(-1,-1,+1)$ & $Z_2$ & $(-1,-1,+1)$ &  $Z_2Z_f$   \\
\hline
$(-1,-1,-1)$ & $X_3$ &   $(-1,-1,-1)$ &  $Z_3Z_f$   & $(-1,-1,-1)$ &  $Z_3Z_f$    &$(-1,-1,-1)$ & $Z_3$  \\
\hline
\end{tabular}
\end{center}
\caption{{\bf Decoding table for the trapped-ion flag-based syndrome extraction:}  (Upper panel) The procedure for the readout of the $S_x^{(p)}$ plaquette stabilizers in order $p=1,2,3$ depends on the outcome of the flag. If no flag is triggered $M_f=+1$, and no error is detected $M_s=+1$ in the syndrome qubit, one can move to the next stabilizer $p+1$. If the flag is not triggered $M_f=+1$, but a syndrome error is detected $M_s=-1$, one proceeds to measure all remaining $X$-type stabilizers with un-flagged circuits, and identify the single-qubit phase-flip error using the leftmost column. If, on the other hand, the flag is triggered $M_f=-1$, one should measure all remaining stabilizers using un-flagged circuits to identify the error, possibly correlated, that has indeed occurred (three remaining columns). The decoding depends on the particular plaquette where the flag is triggered, and is specific to the chosen order,  $p=1,2,3$ in this case.  (Lower panel) The procedure for the readout of the $S_z^{(p)}$ plaquette stabilizers is analogous, but the roles of $Z$ and $X$ are exchanged everywhere. }
\label{table_decoding_flags}
\end{table*}
\begin{table*}
\begin{center}
\begin{tabular}{|c|c|c|c|c|c|c|c|}
\hline
\multicolumn{8}{|c|}{DiVicenzo-Aliferis cat-based readout of the stabilizers }\\
\hline
\multicolumn{4}{|c|}{Correlated error absent $M_{a_3}=-1,M_{a_4}=+1$} &\multicolumn{4}{c|}{  Correlated error present $M_{a_3}=+1,M_{a_4}=-1$}\\
\hline
Syndrome & Error & Syndrome  & Error & Corresponding & Error & Corresponding  & Error \\
 $\left(S_x^{(1)},S_x^{(2)},S_x^{(3)}\right)$ &  correction & $\left(S_z^{(1)},S_z^{(2)},S_z^{(3)}\right)$  &  correction & stabilizer & correction  & stabilizer & correction \\
\hline
 \hspace{2ex}$(+1,+1,+1)$\hspace{2ex} & $\mathbb{I}$ & $\hspace{2ex}(+1,+1,+1)\hspace{2ex}$  & $\mathbb{I}$ & $S_x^{(1)}$ & $Z_3Z_4$ & $S_z^{(1)}$ & $X_3X_4$ \\
\hline
$(+1,+1,-1)$ &\hspace{2ex}  $Z_7$\hspace{2ex} &  $(+1,+1,-1)$& $X_7$&$\hspace{2ex}S_x^{(2)}\hspace{2ex}$ &\hspace{2ex}  $Z_5Z_6$  \hspace{2ex} &$\hspace{2ex} S_z^{(2)}\hspace{2ex}$ & $X_5X_6$  \\
\hline
$(+1,-1,+1)$ &   $Z_5$ &$(+1,-1,+1)$& $X_5$ & $S_x^{(3)}$ &   $Z_6Z_7$   &$S_z^{(3)}$ &\hspace{2ex}  $X_6X_7$ \hspace{2ex} \\
\hline
$(+1,-1,-1)$ &  $Z_6$ &  $(+1,-1,-1)$ &  $X_6$  & & &  &     \\
\hline
$(-1,+1,+1)$ & $Z_1$ &$(-1,+1,+1)$  & $X_1$ &  &  &   &      \\
\hline
$(-1,+1,-1)$ &  $Z_4$ &  $(-1,+1,-1)$ & $X_4$  &  &     &  &      \\
\hline
$(-1,-1,+1)$ & $Z_2$ &   $(-1,-1,+1)$ &  $X_2$  & &  &  &    \\
\hline
$(-1,-1,-1)$ & $Z_3$ &   $(-1,-1,-1)$ &  $X_3$   & &      & &   \\
\hline
\end{tabular}
\end{center}
\caption{{\bf Decoding table for the trapped-ion cat-based syndrome extraction:}  The procedure for the cat-based syndrome extraction of the $S_{\alpha}^{(p)}$ plaquette stabilizers  depends on the combined readout of the pair of ancillary qubits $M_{a_3},M_{a_4}$. If one finds  $M_{a_3}=-1,M_{a_4}=+1$, only a single-qubit error may have occurred at FT level-1, which can be identified from the combined measurements of all the stabilizers according to the two leftmost columns. Conversely,  the values $M_{a_3}=+1,M_{a_4}=-1$ signal that a correlated error may have  propagated into the data qubits. The most-likely dangerous  errors are listed in the two rightmost columns, and depend on which stabilizer was being measured.   }
\label{table_decoding_cat}
\end{table*}

\begin{figure}[t]
 \begin{centering}
  \includegraphics[width=1\columnwidth]{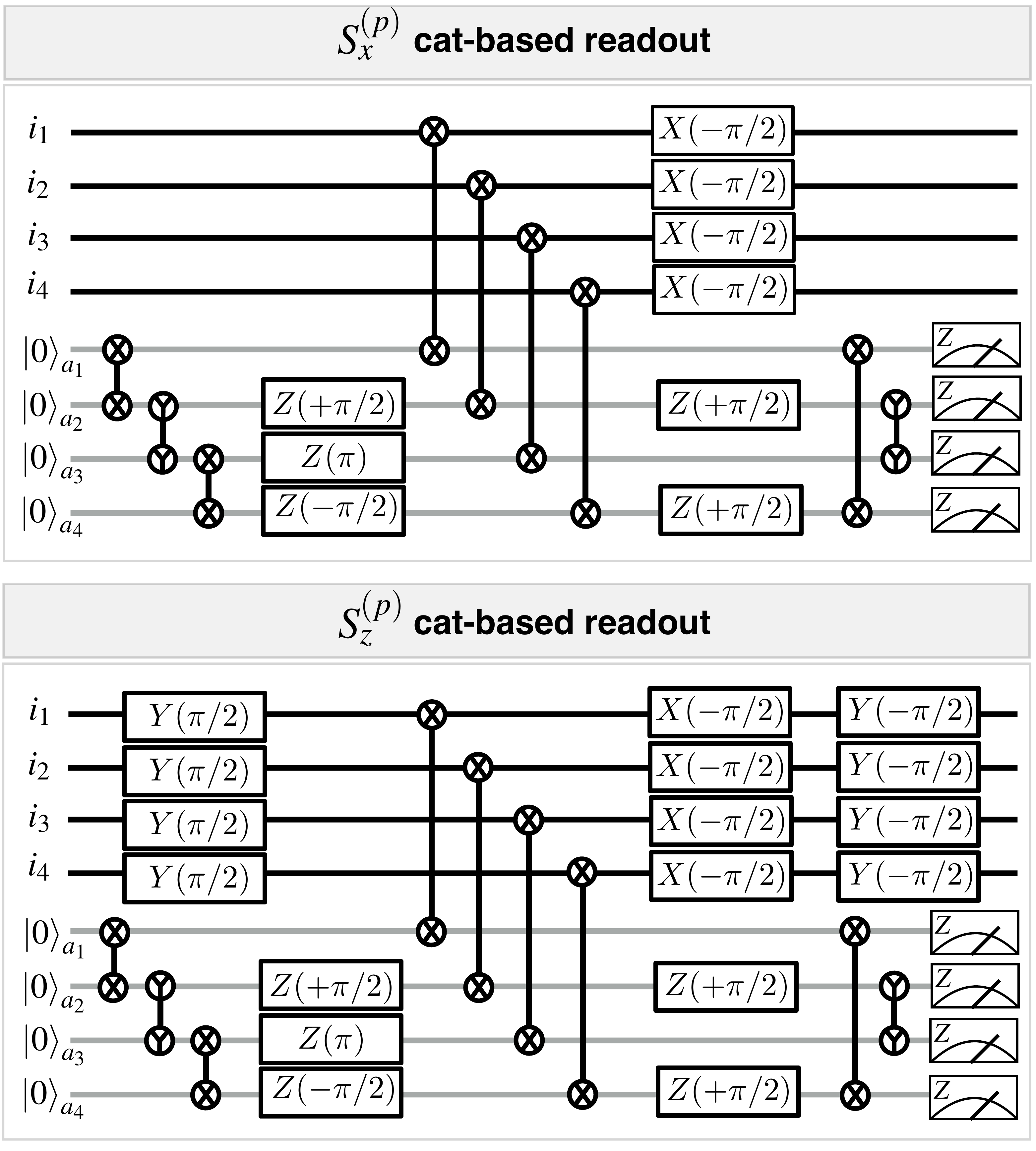}\\
  \caption{\label{Fig:DVA_readout_ms} {\bf Trapped-ion cat-based stabilizer readout:}   Trapped-ion circuit for the  cat-based measurements of $S_x^{(p)},S_z^{(p)}$. The upper qubits  $i_1,i_2,i_3,i_4$, represented by black  lines, correspond to one of the  stabilizers~\eqref{eq:stabilizers} of a particular plaquette (see Fig.~\ref{Fig:7qubitCode}). The lower qubits $a_1,a_2,a_3,a_4$, represented by grey lines, correspond to the syndrome ancillary qubits prepared in a verified cat state from the initial  $\ket{0}$ product state. Here, $a_1,a_2$ contain the stabilizer information, whereas  $a_3,a_4$ are used to verify that no correlated error has occurred during the cat-state preparation, which would cascade into the data qubits. Note again that the $z$-type stabiliser readout is obtained from the $x$-stabiliser readout by a simple $y$-rotation.}
\end{centering}
\end{figure}

Let us now describe the so-called DiVicenzo-Aliferis scheme~\cite{aliferis_ft_qec}, which is the scheme requiring less resources for a trapped-ion implementation within the class of cat-state based approaches~\cite{eQual_1qubit}. In particular,   four ancillary qubits, prepared in a cat-state by a sequence of single- and two-qubit gates, are coupled to the data qubits of a certain plaquette via  sequential MS gates  (see Fig.~\ref{Fig:DVA_readout_ms}). The main idea of this scheme is that the measurements of the ancillary qubits $M_{a_3},M_{a_4}$ can be used to detect if a correlated error may have propagated into the data block, compromising the FT nature of the readout. In particular, if $(M_{a_3},M_{a_4})=(+1,-1)$ during the measurement of the $X$ ($Z$) type stabilizer, the most-likely error is a two-qubit phase (bit) flip error $Z_{i_3}Z_{i_4}$ ($X_{i_3}X_{i_4}$), which must be corrected to guarantee fault tolerance. On the other hand, if $(M_{a_3},M_{a_4})=(-1,+1)$, only a single-qubit error may have occurred, which can be identified by measuring all the remaining stabilizers (see Table~\ref{table_decoding_cat}). We note that the stabilizer information is encoded in the parity of the measurement of the two remaining ancilla qubits $S_\alpha^{(p)}\ket{\Psi} =-M_{a_1}\cdot M_{a_2}\ket{\Psi}$. Let us remark that, to avoid a wrong syndrome extraction due to faulty measurements, the stabilizer readout must be performed twice, or three times if the results do not agree, keeping the syndrome inferred from these last measurements.

It is already apparent by comparing Figs.~\ref{Fig:flag_readout_ions} and~\ref{Fig:DVA_readout_ms} that the cat-based approach  requires more resources than the flag-based scheme, not only in terms of  qubits but also in terms of the required operations. Note also that, although the syndrome extraction of Tables~\ref{table_decoding_flags} and~\ref{table_decoding_cat} seem to be simpler for the cat-based approach, the cat-based approach indeed requires more resources in terms of operations, as the readout needs to be performed up to three times to discard wrong syndromes due to measurement errors. We also note that  the QCCD trapped-ion implementation of these circuits will contain a considerable overhead of the other elementary operations of Table~\ref{tab:summary_toolbox}, such as  various required crystal reconfigurations. Therefore, the resource-consuming nature of the cat-based approach can only get amplified when one considers a realistic trapped-ion implementation. The goal of this work is to explore how these differences affect the  QEC performance at a quantitative level, and determine if the flag-based approach can be an important improvement for the demonstration of beneficial QEC in trapped ions under a  realistic microscopic  model of noise.

\subsection{Optimized FT flag-based encoding}

 \begin{figure*}[t]
 \begin{centering}
  \includegraphics[width=1.8\columnwidth]{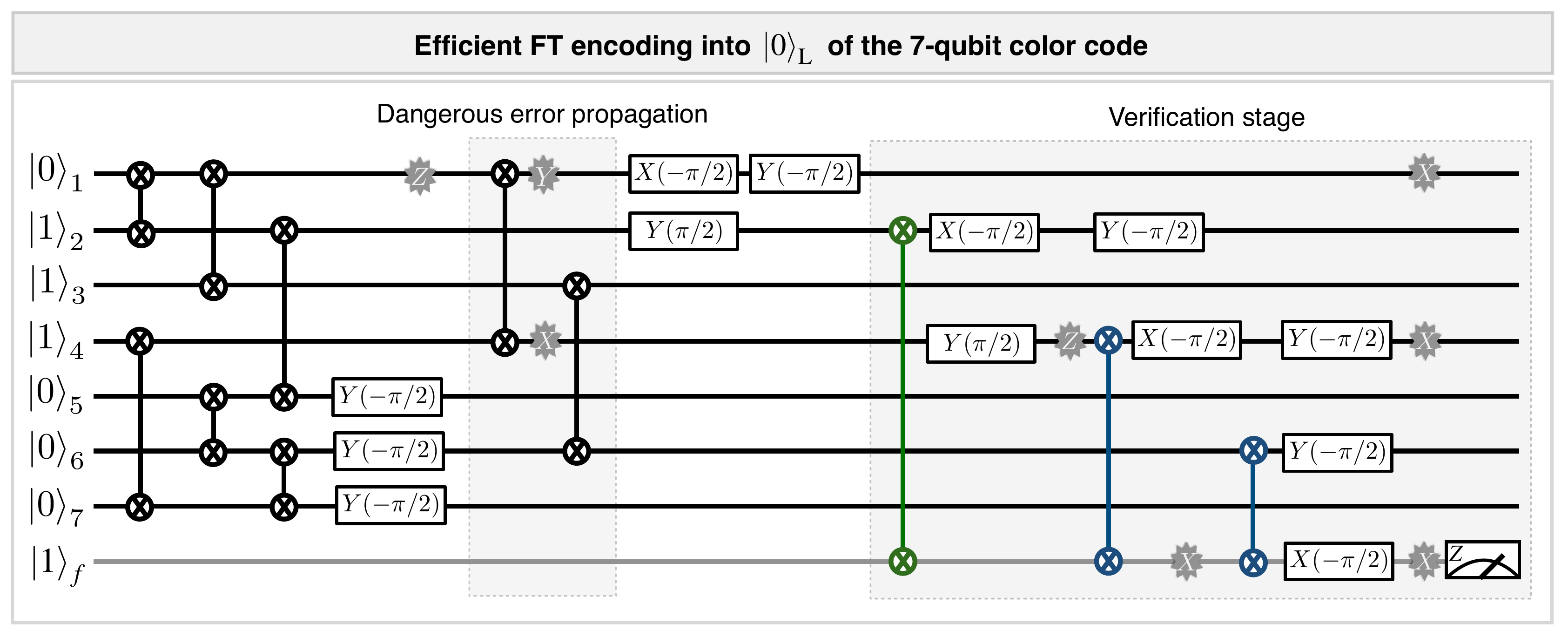}\\
  \caption{\label{Fig:efficient_encoding_ft_ms}  {\bf Fault-tolerant MS-based encoding in the 7-qubit color code:} Encoding circuit for the logical $\ket{0}_{\rm L}$  based on a particular sequence of 8 MS gates, and 6 single-qubit rotations. The last pair of MS gates, inside a shaded rectangle,  can lead to two-qubit  errors in the data block, and must be detected by a flag qubit to ensure fault-tolerance. This verification step, inside the rightmost shaded rectangle, consists of 3 MS gates that couple data qubits to the ancillary qubit (in grey), and 7 single-qubit rotations. By using the error-propagation of MS gates mentioned above, one can check that the single phase-flip error during the preparation turns into a pair of bit-flip errors, detected by the flag measurement $M_{f}=-1$. The corresponding errors are represented with grey stars, and the labels indicate their type as one propagates through the circuit.  }
\end{centering}
\end{figure*}

Another important   QEC operation is the redundant encoding of a particular logical state, such as $\ket{0}_{\rm L}$ or ${\ket{+}_{\rm L}}$. An advantage of the stabilizer formalism   is the existence of  clear  strategies to accomplish such encoding~\cite{nielsen-book}. For the 7-qubit color code, one may start  from $\ket{\psi_0}=\otimes_{i}\ket{0}_i$, and measure fault-tolerantly all $X$-type stabilizers~\eqref{eq:stabilizers} following, for instance,  the two approaches  discussed above. Depending on the outcome of these measurements, one can perform the corresponding operations to project into the code subspace, e.g. into $\ket{0}_{\rm L}$. Likewise, for the   $\ket{+}_{\rm L}$ encoding, one simply needs to exchange the roles of the $x$ and $z$ bases. Note however, that this is a resource-intensive  approach, as it requires performing full rounds of active QEC.

Interestingly,   there are more efficient encoding strategies  based on  a verification step to ascertain that the desired level of fault-tolerance has been achieved.  Starting from Steane's encoding based on  Latin rectangles~\cite{steane_squares}, it is possible to optimize  the encoding   into  $\ket{0}_{\rm L}$ following  the strategy discussed in~\cite{Reichhardt_encoding}. However, if errors  take place, the imperfect encoding becomes non-FT, as pairs of data qubits from the same logical block are coupled, and  single errors can cascade into multiple  errors. We now describe  a flag-based approach that can detect and correct such dangerous correlated errors (see Fig.~\ref{Fig:efficient_encoding_ft_ms}).

 As realized by Goto~\cite{Goto2016} for the CNOT version of this circuit,  prior to the last pair of entangling gates inside the leftmost shaded rectangle of Fig.~\ref{Fig:efficient_encoding_ft_ms}, all propagated errors are equivalent to  single-qubit errors. Since these  are correctable within the 7-qubit color code,  one can  ensure fault-tolerance at level-1 by making sure that no correlated error is being created by the last pair of MS gates.  The verification step can be thus greatly simplified, as it only requires detecting two dangerous bit flips, one of which is highlighted in Fig.~\ref{Fig:efficient_encoding_ft_ms}. The verification can be accomplished by two  additional MS gates  that couple the data qubits to an  ancillary flag qubit (see the blue pair of gates in Fig.~\ref{Fig:efficient_encoding_ft_ms}). As depicted in this figure, one can detect when such a correlated error has occurred by measuring the flag qubit in the $z$-basis $M_f=-1$. Additionally, by  introducing an MS gate between the first data qubit and the ancilla (see the green gate in Fig.~\ref{Fig:efficient_encoding_ft_ms}), the measurement  gives information about the logical operator of the 7-qubit code, which indeed stabilizes the logical $\ket{0}_{\rm L}$ state, such that  the target encoding into $\ket{0}_{\rm L}$ is not altered by the verification step.

\begin{figure}[t]
 \begin{centering}
  \includegraphics[width=1.0\columnwidth]{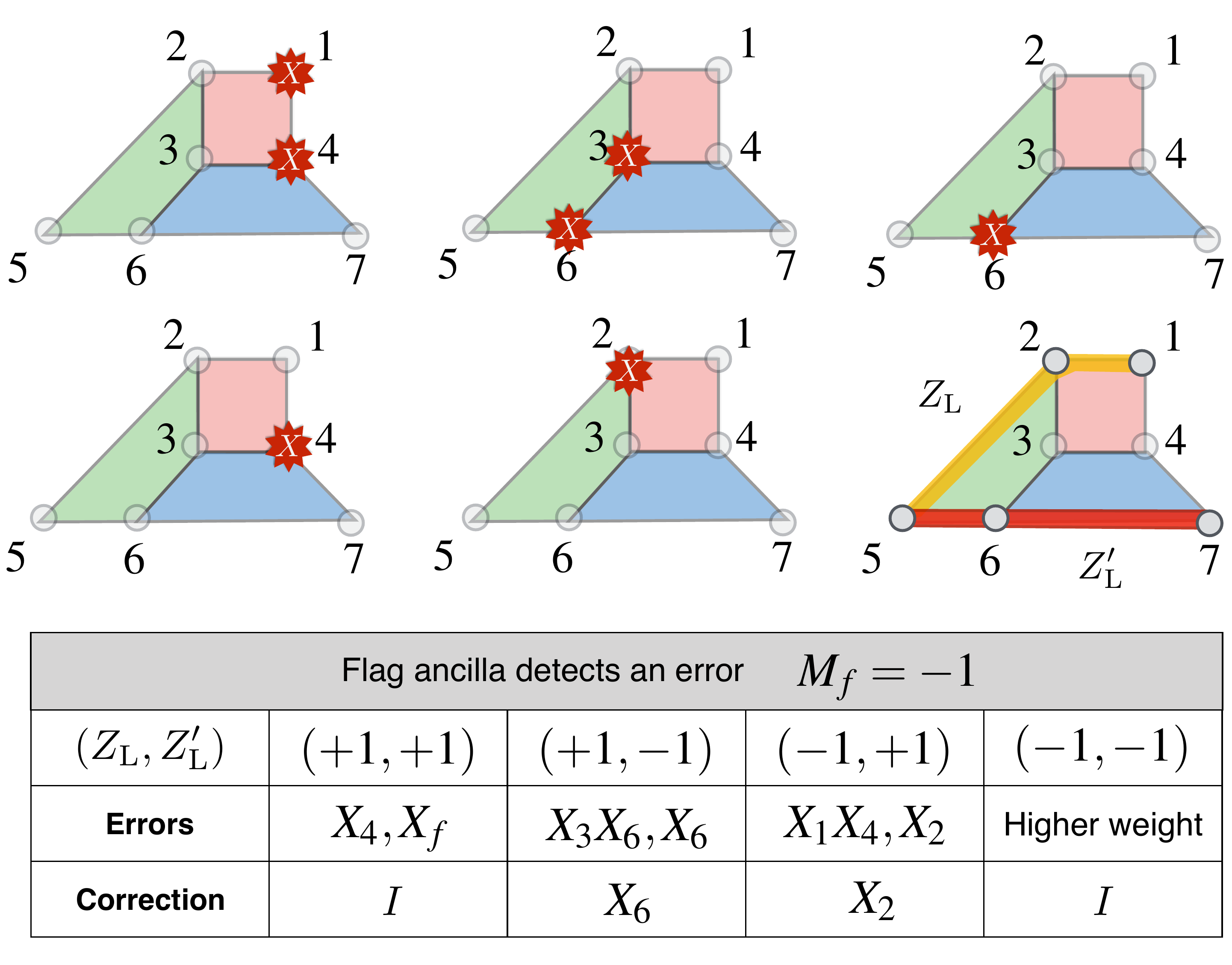}\\
  \caption{\label{Fig:error_correction_encoding}  {\bf Flag-based correction of encoding correlated errors:} (Upper panel) Possible errors, represented by red stars, which may have taken place when the ancilla measurement of Fig.~\ref{Fig:efficient_encoding_ft_ms} yields $M_f=-1$. By measuring the two equivalent  logical operators $Z_{\rm L}$ and $Z_{\rm L}'$, it is possible to find single-qubit corrections that will turn the possible errors  into single-qubit errors (lower panel) that can be corrected by the QEC code.   Note that when the logical outcomes are $Z_{\rm L}=Z_{\rm L}'=-1$, this should be caused by an operator that involves more qubits and has not arisen via propagation, but instead by multiple errors occurring on the bare physical qubits. These errors cannot be corrected by the code anyway, so the flag-based routine is FT at level-1. }
\end{centering}
\end{figure}

 Let us now go beyond by showing that one can, not only detect when a correlated error may have occurred and use post-selection to achieve a FT encoding, but also distinguish between the two types of correlated errors, and correct them  instead of using post-selection. The philosophy is similar to the flag-based approach, as it relies on   additional measurements, which can be used in combination with the flag measurement to correct for the dangerous correlated errors. By measuring the logical operators  $Z_{\rm L}=Z_{1}Z_{2}Z_{5}$ and $Z'_{\rm L}=Z_{5}Z_{6}Z_{7}$ of Fig.~\ref{Fig:error_correction_encoding}, we can ascertain that the two-qubit errors become equivalent to single-qubit errors (up to the code stabilisers) after the corrections  listed in the lower table of Fig.~\ref{Fig:error_correction_encoding}, such that fault tolerance at level-1 is achieved without any post-selection. Let us finally note that the FT encoding into the logical  $\ket{+}_{\rm L}$ state can be achieved using the same circuit, and applying a Hadamard gate (see Fig.~\ref{Fig:gate_set}) to all  qubits in the data block, right at the end of the circuit. This is a direct consequence of the transversality of the Hadamard gate in color codes.

\section{\bf Numerical assessment of the FT performance of trapped-ion QEC}
\label{sec:numerical_results}

\subsection{Assessing the benefit of trapped-ion QEC}

 A possible criterion to determine the success of  FT-QEC  would be to demonstrate a reduced error rate for a set of representative quantum circuits, including non-Clifford operations,  with respect to the best-possible un-encoded qubits~\cite{gottesman_small_codes}. This criterion, however, is  too demanding for near-future experiments in various technologies. For trapped ions, the very large coherence times on the order of seconds, together with the very high fidelities for elementary operations exceeding $99.9\%$,  indicate that the above criterion will be extremely difficult to meet with near-term devices considering the overhead in  complexity  for QEC (i.e. these bare operations/qualities are already so good that it is unlikely that short-distance QEC codes, with an increased complexity and making use of lower-fidelity entangling gates, will allow to improve them further). In order to assess the progress of near-term QEC, a milder criterion would be to demonstrate that a complete round of error detection and correction proves to be beneficial rather than detrimental. This criterion should be applied to a setup that, although it might not use the best bare qubits nor the best single- and two-qubit  gates, it suffices to accommodate all the requirements for QEC with small codes, and has a clear route for scaling up towards larger distances. 
 
 This criterion can be translated into a quantitative, and experimentally relevant, assessment to guide future quantum hardware developments by introducing the concept of integrity of a quantum memory for a particular quantum-information task: state discrimination~\cite{eQual_1qubit}. As formalized in~\cite{AIBpreprint}, we start by considering that a qubit with density matrix $\rho$ is to be stored in a code-based memory channel $\Phi$  for a specified period of time through the following process. {\it (i) Setup:} At $t=0$ Alice  encodes the single qubit $\rho$ into an $n$-qubit logical code $\rho_n=\mathsf E(\rho)$ where $\mathsf E$ is the encoding map.
{\it (ii) Noise channel:} Evolution and degradation of the logical qubit occurs while it is stored, and which may include the effects of active QEC, informally denoted as action by Igor.
  We have $\rho_n^\prime=\mathsf N(\rho_n)$ where $\mathsf N$ is the noise map.
 {\it (iii) Conclusion:} At $t=\tau$, Bob  performs a QEC cycle, and then reverses Alice's encoding process to obtain a single physical qubit $\tilde\rho=\mathsf D(\rho^\prime_n)$ where $\mathsf D$ is the decoding map.

 We want to evaluate quantitatively the integrity of the entire channel $\Phi = {\mathsf D \circ \mathsf N \circ \mathsf E}$, thus incorporating Alice's encoding $\mathsf E$, the noise and Igor's QEC $\mathsf N$, and Bob's decoding $\mathsf D$. Intuitively, integrity is ``the probability that Bob, receiving a logical qubit from the memory, can infer its state''. More precisely, it is the probability that Bob can still distinguish two states $\psi$, $\psi_\perp$, that were, when originally prepared by Alice, fully distinguishable~\cite{AIBpreprint}. Thus, the integrity $\mathcal R$ of the quantum memory channel can be defined as 
\begin{equation}
 \ \   \mathcal R(\Phi) =
    \min_{\psi}
      \mathcal D\bigl(\Phi(\psi), \Phi(\psi_\perp)\bigr),\ \ 
      \label{eqn:integrityShortDef}
\end{equation}
where $\mathcal D(A,B)=\half{\rm Tr}\left\{((A-B)^\dagger(A-B))^{1/2}\right\}$ is the trace distance derived from the trace norm~\cite{nielsen-book}. We recall that the maximum probability that an ideal experimentalist would succeed in identifying which of two states, $\rho_1$ and $\rho_2$, he/she has been presented with, given a $50/50$ prior, is
$
p_{g}=\tfrac{1}{2}+\tfrac{1}{2}  \mathcal R(\rho_1,\rho_2).
$
Therefore, the minimum in definition (\ref{eqn:integrityShortDef}) ensures that we consider all possible states that Alice might encode,  reporting the worst-possible degradation.
To measure the integrity of a memory channel, we consider that Bob is given the information that Alice's initial qubit was $\psi$ or  $\psi_\perp$. Bob then makes a  guess by a measurement  of his choice to determine if $\Phi(\rho)$ is $\Phi(\psi)$ or $\Phi(\psi_\perp)$, with worst-case probability  $p_g$, such that  $\mathcal R(\Phi)=2p_{g}-1.$  A key enabling observation~\cite{AIBpreprint} is that, for a wide variety of error models,  the minimum of Eq.~(\ref{eqn:integrityShortDef}) can be found  by having Alice only preparing logical states in the Pauli $x$-, $y$- or $z$-basis. Moreover, Bob will have optimal performance when he measures in the same basis which Alice has used, which makes the protocol experimentally practical.

The integrity can be used to quantify the beneficial nature of QEC (i.e. the level of success of Igor)  by using  $\Phi^m$ to label the memory channel when Igor performs $m$ rounds of error correction~\cite{AIBpreprint}, and defining a series of milestones:

{\it (M1) Beneficial error correction}:  A round of QEC is beneficial if Bob's probability of subsequently discriminating the state correctly is higher when Igor indeed performs that round 
\begin{equation}
\mathcal R(\Phi^1) >  \mathcal R(\Phi^0)\ \ \ \text{for some memory time }\tau.  
\label{eq:IgorIsGood}
\end{equation}

{\it (M2) Beneficial multi-round error correction}:
For a sufficiently high performing Igor, and  a long $\tau$, it will be beneficial to have multiple rounds of correction.  This will be a signature of  progress toward a practical quantum memory
\begin{equation}
 \mathcal R(\Phi^m) >  \mathcal R(\Phi^{m-1}) \ \ \ \text{for some memory time  }\tau.  
\label{eq:nRounds}
\end{equation}

These two milestones compare the memory integrity of an error-corrected encoded qubit with an encoded qubit without additional QEC. However, we would also like to 
address the question of whether it is worth using encoded memories at all, which requires comparing to  error-corrected encoded qubit with  a simple un-encoded memory. We use the symbol $\Theta$ for that memory channel, and define the following milestones:

{\it (M3) Beneficial encoded memory}: The actively-corrected encoded memory  beats the simple single-qubit memory
\begin{equation}
\label{eq:versusSingleQubitMem}
\mathcal{R}(\Phi^m)>\mathcal{R}(\Theta)\ \ \text{for some memory time  }\tau.
\end{equation}
Here we require only that this is true for some specific value of $m>0$. We note that  this milestone and the concept of a `pseudo-threshold'~\cite{svore2006,cross2009} can be related~\cite{AIBpreprint}.

{\it (M4) Strictly superior encoded memory}: 
The most challenging goal of our memory characterization is 
\begin{equation}
\label{eq:unconditionallyBetter}
\max_m\mathcal{R}(\Phi^m)>\mathcal{R}(\Theta)\ \ \text{for any memory time  }\tau. 
\end{equation}
Here, the maximum is over a family of memory channels having the same duration, but with differing numbers of error correction cycles $m$. If this condition is satisfied, it means that for any desired duration $\tau$, we can sustain our encoded quantum memory at a higher integrity than a single physical qubit memory by applying a suitable number of QEC cycles.

Building on these milestones, we describe in the following sections numerical results obtained from full wave-function simulations, which aim at assessing the   performance of the trapped-ion color code under the improved microscopic error models. These numerical results replace the previous study with a simpler noise model~\cite{eQual_1qubit}, which furthermore assumed ideal non-faulty encoding $\mathsf{E}$ and decoding $\mathsf{D}$ maps. In this way, the present results represent a non-trivial increase in sophistication of the description of the experimental architecture and, we believe, one of the most sophisticated numerical simulations of QEC in  a realistic quantum processor to date. 

\subsection{Leakage noise:  beneficial  QEC by repumping}
\label{subsec:numericsForTrueNoise}

Let us start by exploring numerically the effects of leakage noise since, as argued above, it has the potential of causing a large detrimental effect on the performance of trapped-ion QEC. We recall that the 7-qubit color code can correct for bit- and phase-flip errors, but cannot overcome the effects of  population  leaked from the computational subspace. As discussed in the previous section, the Alice-Igor-Bob scenario of state discrimination offers a clear, intuitive, and  quantitative method to assess the prospects of trapped-ion hardware to demonstrate  beneficial  QEC~\cite{eQual_1qubit,AIBpreprint}. In a previous study~\cite{eQual_1qubit}, in order to single out Igor's  QEC capabilities clearly,  Alice and Bob were modeled as ideal agents that can encode and decode any quantum state perfectly. However,  Alice's encoding and Bob's decoding will present imperfections in any realistic experiment, which could  interfere and complicate our assessment of the beneficial role of Igor. In this section, we quantify this potential interference  by numerically studying the memory integrity  using faulty encoding/decoding strategies.

\begin{figure}[t]
 \begin{centering}
  \includegraphics[width=1.0\columnwidth]{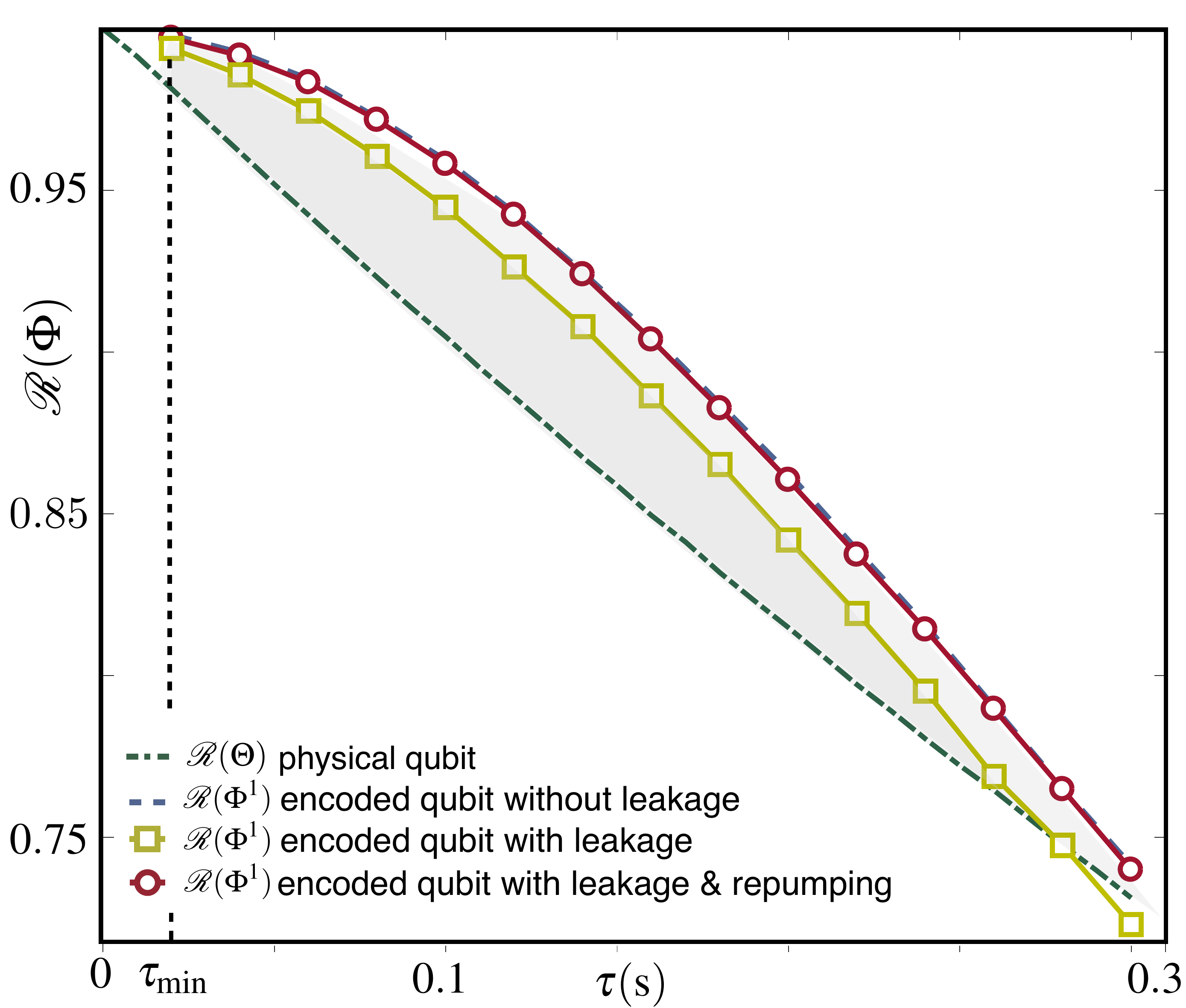}\\
  \caption{\label{Fig:error_correction_rempumping_results}  {\bf Beneficial QEC with repumping sequence:} Integrity of bare and encoded quantum memories as a function of the memory time $\tau$. The green dashed-dotted line represents the integrity degradation for a bare physical qubit due to environmental noise, including both dephasing and spontaneous emission including a 4/9 damping/leakage ratio. The yellow squares represent   the integrity for an encoded memory when Igor performs a single round of QEC at $\tau/2$, which aims to correct errors from the environmental noise and Alice's imperfect encoding. When the repumping sequence is applied prior to gates of Igor's QEC cycle, the memory integrity increases considerably (red circles), almost  reaching  the maximum set by a noise model where spontaneous emission only occurs in the damping channel without any leakage (blue dashed line).      }
\end{centering}
\end{figure}

To understand the  impact of leakage, we  use the aforementioned Alice-Igor-Bob framework, where Alice encodes imperfectly a logical $|+\rangle_L$ (or $|-\rangle_L$), which then experiences a period of environmental noise  before Igor does a round of imperfect flag-based QEC. Finally, this logical state is subjected to a second period of environmental exposure, before an imperfect 
Bob finally attempts to determine whether Alice created  $|+\rangle_L$ or  $|-\rangle_L$.  For the environmental noise model, we consider the improved microscopic error model of Sec.~\ref{sec:noise_models}, and artificially switch on/off the leakage and the repumping sequence, which will be applied prior to gates of Igor's attempt at QEC. In Fig.~\ref{Fig:error_correction_rempumping_results}, we present these numerical studies, which will allow us to discuss neatly the dangerous effects of leakage, and how to combat them with repumping.

The green dashed-dotted line represents the integrity $\mathcal{R}(\Theta)$ of the  un-encoded memory using a bare physical qubit, and only serves the purpose of providing a guide-to-the-eye for  the expected degradation of the memory due to the trapped-ion environmental noise, which includes damping and leakage. Let us note that the linear decay of the integrity  display, in this case,  how  single-qubit errors acting on the physical qubit affect the bare physical memory. The blue dashed line stands for  the integrity of the  encoded memory, after a single round of flag-based QEC by Igor $\mathcal{R}(\Phi^1)$, where the spontaneous emission from the metastable state only contributes with amplitude damping (i.e.~we artificially set $\Gamma'=0, \Gamma=1/T_1$, thus switching off the leakage). In this way, this line serves as a guide-to-the-eye for the optimal beneficial effects of Igor's QEC, as the single-qubit environmental noise and gate errors can now be corrected by the 7-qubit color code. We note that the characteristic non-linear decay of the integrity gives  direct evidence of the fault-tolerance of the scheme (i.e. only two or more errors can affect the encoded memory, such that the short-time scaling of the integrity is quadratic rather than linear). The same interpretation of the characteristic scaling applies to other figures in the manuscript.  Notice how the memory integrity is only defined after a finite $\tau_{\rm min}$, which corresponds to the time required by Igor's flag-based QEC cycle. For $\tau>\tau_{\rm min}$, the encoded quantum memory is subjected to additional environmental noise in the time lapses before and after Igor's attempt at QEC. Let us also note that, in contrast to the bare memory, the encoded memory integrity would not not start from the maximal value $\mathcal{R}(\Phi^1)=1$ even for a perfect encoding by Alice. Instead,  it does start from a lower integrity, as the imperfect QEC contributes  to the initial degradation. The clear advantage of QEC is that the initial slope of the integrity signals a slower degradation with respect to the bare single-qubit memory.    

With these limiting cases, we can now understand the effects of leakage noise.
The yellow squares represent the  same QEC integrity $\mathcal{R}(\Phi^1)$, but this time activating a maximal leakage  (i.e. $\Gamma=\Gamma'=1/2T_1$). As the figure clearly shows, the integrity gets degraded considerably faster with respect to the noise model without leakage (blue dashed line), leading to a much smaller region where QEC is beneficial in comparison to the bare single-qubit memory (i.e. compare the shaded grey regions). This confirms our previous expectation that the effect of leakage can have important detrimental effects and compromise considerably the prospects of demonstrating beneficial QEC in near-term architectures. Let us note that, although this limiting case $\Gamma=\Gamma'$ does not represent the optical qubit, it gives a qualitative description of the maximal effect of leakage in other qubit choices where the leakage can be higher (i.e. hyperfine qubits driven by two-photon Raman transitions~\cite{leakage_hyperfine} ). Let us now discuss how the repumping scheme can overcome the additional leakage degradation even in this  worst-case scenario. As shown by the red circles,   which represent the  integrity $\mathcal{R}(\Phi^1)$ when  repumping is applied prior to gates of the QEC cycle, the degradation of the integrity almost reaches the optimal case where all the spontaneous decay occurs in the amplitude damping channel (i.e. blue dashed line). These results also confirm our previous statement that the repumping  turns leakage into a sort of  amplitude damping, which is correctable by the QEC code.

\subsection{Break-even point: Assessing the performance of flag- and cat-based trapped-ion QEC}
\label{subsec:numericsForFlag}

Once the method to combat leakage noise has been benchmarked with our microscopic model of flag-based trapped-ion QEC, let us move on to  a comparative numerical study of the performance of cat- and flag-based approaches for QEC with the trapped-ion 7-qubit color code. We now use the complete and realistic microscopic noise model discussed at length in Sec.~\ref{sec:noise_models}, and set the leakage and amplitude-damping rates to the experimental value of $\Gamma'=4\Gamma/9$. We remind the reader that, in order to simulate numerically the different QEC approaches, we have to translate the corresponding circuits (see Figs.~\ref{Fig:flag_readout_ions},~\ref{Fig:DVA_readout_ms} and~\ref{Fig:error_correction_encoding}) into the corresponding microscopic schedules with the sequence of elementary operations in the QCCD trapped-ion processor (see Appendix~\eqref{sec:appendix_B}). These microscopic schedules are then translated into the corresponding sequence of faulty operations and idle periods where the environmental noise affects the idle qubits, which is then numerically simulated using our full wave-function Monte Carlo approach.

\begin{figure*}[t]
 \begin{centering}
  \includegraphics[width=2.0\columnwidth]{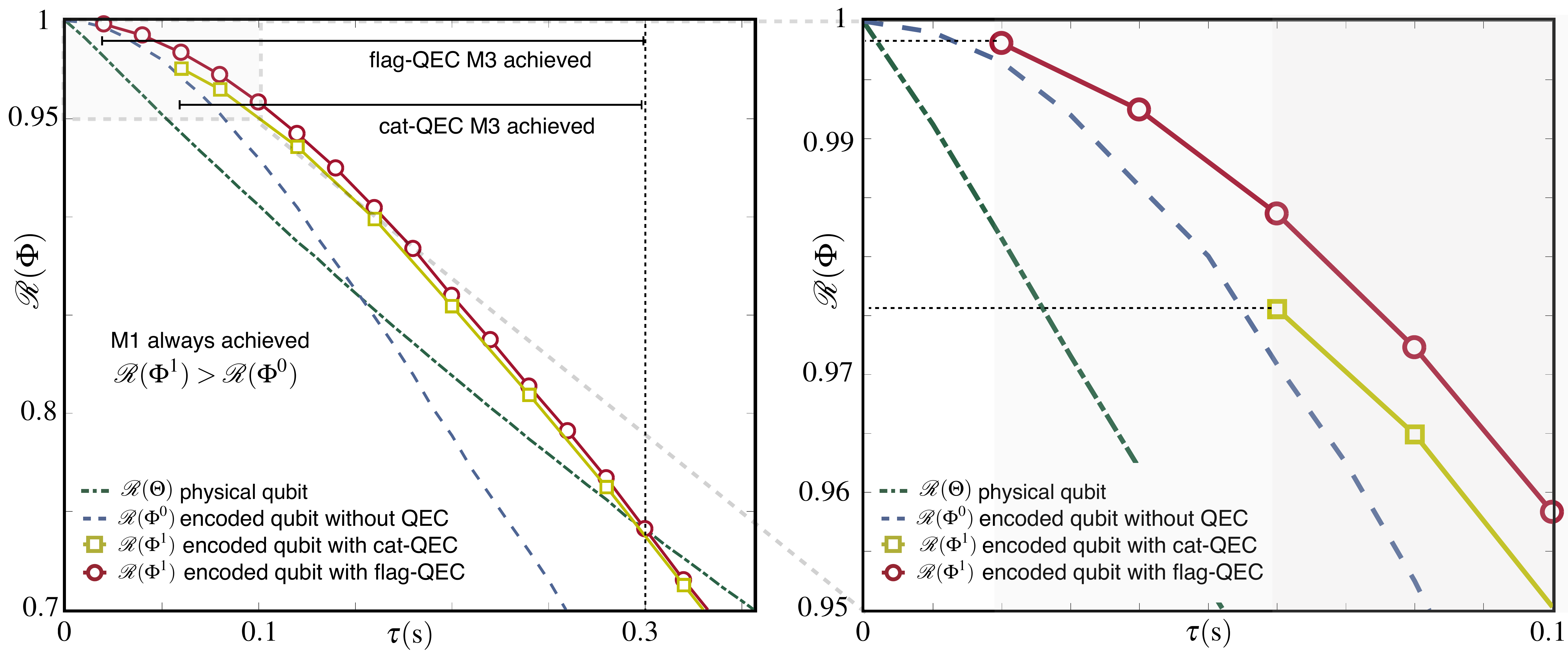}\\
  \caption{\label{Fig:cat_flag_results}  {\bf Comparative assessment of cat- and flag-based trapped-ion QEC:} Integrity of bare and encoded quantum memories as a function of the memory time $\tau$. The green dashed-dotted line represents the integrity degradation for a bare physical qubit, the blue dashed line stands for the integrity of an encoded but un-corrected memory, and the yellow squares and red circles represent the memory integrity for a 7-qubit color code corrected via a single round of cat- and flag-based QEC, respectively. The right panel shows a short-time inset of the integrity degradation of the various memories, and allows for a clearer comparison between the cat- and flag-based approaches.    }
\end{centering}
\end{figure*}

Let us start by considering the milestones {M1} and {M3}, which will allow us to explore the beneficial nature of a single round of QEC, comparing the potential of cat- and flag-based approaches. The numerical results for the Alice-Igor-Bob scheme to quantify the memory integrity of both QEC schemes are presented in Fig.~\ref{Fig:cat_flag_results}. In the left panel, we use the green dashed-dotted (blue dashed) line as a  guide-to-the-eye representing the integrity degradation of a bare physical (encoded but not corrected) memory. In addition, we represent the encoded memory integrity when Igor performs a single round of QEC using the cat-based (yellow squares) and flag-based (red circles) approaches. This figure clearly shows that, for memory times beyond the corresponding $\tau_{\rm min}$, both  schemes are always better than the encoded but not QEC memory (blue dashed line), thus meeting the criterion~\eqref{eq:IgorIsGood} for milestone M1 of beneficial error correction. Let us note that this is in general not always guaranteed, as the attempt of QEC introduces a large overhead of additional faulty operations that may introduce more noise than the one that the code can correct for. One can see from the inset of the right panel that the effect of the faulty QEC is to start off from a lower integrity (see the dashed black lines), which could potentially lie below the un-corrected encoded integrity (blue dashed line) if Igor's QEC capabilities were not sufficiently good. In the present case, however, our numerical results show that the expected near-term improvements of trapped-ion hardware (see Table~\ref{tab:summary_toolbox}) can suffice to demonstrate beneficial QEC M1 using either of the cat- or flag-based approaches. 

Let us note, however, that the more resource-intensive cat-based approach performs considerably worse than the flag-based approach (see the inset of the right panel for small memory times). In this figure, one observes that the minimal time $\tau_{\rm min}$ of the cat-based approach is much larger than that of the flag-based QEC, which is a consequence of the larger depth of the circuits, and higher number of crystal reconfiguration operations leading to longer idle periods, which are required to prepare the cat states ensuring FT at level-1. As a consequence of this larger depth, the cat-based encoded memory is exposed to more errors and environmental noise, such that the corresponding integrity (first yellow square) starts off at a considerably lower value with respect to the flag-based approach (first red circle). This flag-based improvement will be important when we consider multiple rounds of QEC, as the high short-time integrity can then be maintained for larger and larger memory times with a very small degradation.

Let us now address the milestone M3~\eqref{eq:versusSingleQubitMem}, which aims at quantifying a beneficial encoded memory when the encoded and error-corrected memory performs better than   the bare physical qubit. As depicted in the left panel of Fig.~\ref{Fig:cat_flag_results}, this milestone can be achieved again for both cat- and flag-based approaches with the expected trapped-ion resources. This milestone can be achieved for a wide range of memory times  $\tau\in[\tau_{\rm min},\tau_{\rm max}]$, which cannot be either too short nor too large. At very short times, the unencoded qubit will always beat the QEC memory as Igor does not have sufficient time to perform the full round of QEC. On the other hand, at very large times, the environmental noise keeps affecting the multi-qubit memory introducing errors that can no longer be corrected, and degrades the integrity beyond that of a single un-encoded qubit. Let us note that, once again, the region  $\tau\in[\tau_{\rm min},\tau_{\rm max}]$ where M3 is achieved is considerably larger for the flag-based approach than for the cat-based one.

\begin{figure}[t]
 \begin{centering}
  \includegraphics[width=1.0\columnwidth]{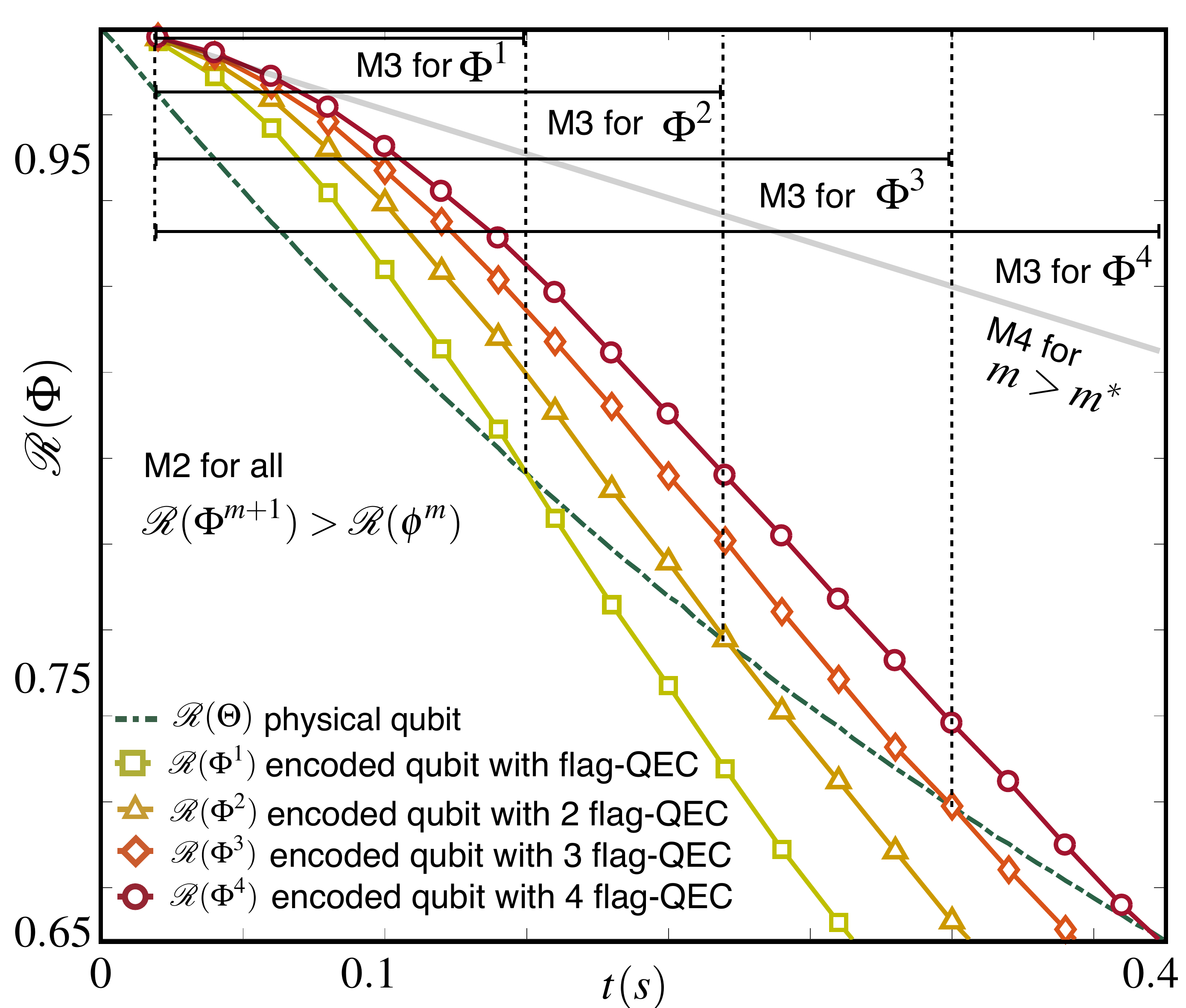}\\
  \caption{\label{Fig:multiple_flag_results}  {\bf Beneficial QEC with multiple rounds of QEC:} Integrity of bare and encoded quantum memories as a function of the memory time $\tau$. The green dashed-dotted line represents the integrity degradation for a bare physical qubit,  while the symbols stand for the integrity of error-corrected 7-qubit color-code memories which are subjected to $m$ rounds of flag-based QEC (i.e. $m=1$ yellow squares, $m=2$ orange triangles, $m=3$ red diamonds, and $m=4$ dark-red circles). We also display the regions where various milestones can be achieved, as well as the asymptotic slope for frequent QEC rounds.  }
\end{centering}
\end{figure}

Having concluded that the flag-based approach to trapped-ion QEC yields a clear advantage, let us now address the milestones M2 and M4 for an encoded quantum memory with multiple rounds of flag-based QEC (see Fig.~\ref{Fig:multiple_flag_results}). In this figure we represent the integrity degradation for the 7-qubit color code quantum memory after a time $\tau$, where a number of non-perfect  flag-based QEC cycles are interspersed between periods of pure environmental noise (i.e.~yellow squares $m=1$ round of QEC $\mathcal{R}(\Phi^1)$, orange triangles  $m=2$ rounds of QEC $\mathcal{R}(\Phi^2)$, red  diamonds $m=3$ rounds of QEC $\mathcal{R}(\Phi^3)$, and dark-red circles   $m=4$ rounds of QEC $\mathcal{R}(\Phi^4)$). We note that, prior to each of Igor's attempts at QEC, we apply a repumping sequence to project the leaked population back to the computational subspace. In this figure, the green dashed line represents, again, the integrity degradation of a bare un-encoded memory, which sets the standard that the QEC memory must beat to achieve the milestones.

Regarding milestone M2 for  beneficial multi-round error correction, we note that the corresponding criterion~\eqref{eq:nRounds} is clearly met for all of the displayed memory times, as the integrity of the memory is readily improved as more rounds of QEC are applied. This figure also shows that milestone M3 for a beneficial encoded memory beating the bare physical qubit~\eqref{eq:versusSingleQubitMem} is achieved for larger and larger memory times as one increases the number of rounds of Igor's QEC. Finally, regarding the final and more stringent milestone M4 for a strictly superior encoded memory, we recall that the corresponding criterion~\eqref{eq:unconditionallyBetter} requires that it is always possible to beat the bare memory for any target memory duration by applying sufficiently many rounds of QEC. In Fig.~\ref{Fig:multiple_flag_results}, one can identify a  trend of the memory integrity as one increases the number of QEC cycles, whereby the slow-time QEC protection can be extended to longer and longer memory times as more rounds of QEC are applied. In the limit of many cycles, we depict a qualitative straight envelope with a  slope that improves the robustness of the bare physical memory. Accordingly, our numerical simulations suggest    that it should be possible to beat the bare memory for any target $\tau$. For instance, if the target memory time is $\tau\leq0.4s$, our numerical results show that it suffices to apply $m=4$ rounds of flag-based QEC to beat the bare memory.

\section{\bf Conclusions and Outlook}
\label{sec:conclusions}

We have presented a detailed account of QEC in near-term trapped-ion devices, considering important aspects that complement and improve the recent study of~\cite{eQual_1qubit}. Firstly, we have improved  the previous microscopic error model in several directions: in the present work we consider coherent and correlated noise, non-Pauli errors such as amplitude damping and qubit leakage, and a more refined microscopic error model for the entangling MS gates. With these improvements, we believe that our  current description of  trapped-ion QEC in segmented dual-species ion traps contains one of the most realistic error models in QEC studies to date.  Secondly, we have discussed the trapped-ion implementation of a new set of FT QEC tools based on the use of flag qubits, both for active detection and correction of errors but also for a FT optimized encoding. These flag-based trapped-ion QEC tools rely on a realistic modeling of the hardware capabilities and microscopic schedules that underlie the more abstract circuit-based approach, and  have the potential to  change the prospects of experimental trapped-ion QEC with near-term devices. Finally, we have presented detailed Monte Carlo full-wavefunction numerical simulations to assess the QEC capabilities of this flag-based approach, and to compare it to other more resource-intensive FT schemes. Our  simulations which, to date, may constitute the most sophisticated numerical account of QEC  under a complex non-Pauli noise model, clearly show that the flag-based approach is superior for the expected trapped-ion technologies. This statement is substantiated quantitatively by comparing the memory integrity of the QEC codes, and its potential to achieve various milestones in an experimentally relevant scenario.  We believe that this study will be useful to guide near-term efforts for QEC with trapped-ion quantum processors.

\acknowledgements

The research is based upon work supported by the Office of the Director of National Intelligence (ODNI), Intelligence Advanced Research Projects Activity (IARPA), via the U.S. Army Research Office Grant No. W911NF-16-1- 0070. The views and conclusions contained herein are those of the authors and should not be interpreted as necessarily representing the official policies or endorsements, either expressed or implied, of the ODNI, IARPA, or the U.S. Government. The U.S. Government is authorized to reproduce and distribute reprints for Governmental purposes notwithstanding any copyright annotation thereon. Any opinions, findings, and conclusions or recommendations expressed in this mate- rial are those of the author(s) and do not necessarily reflect the view of the U.S. Army Research Office.

We also acknowledge support by U.S. A.R.O. through Grant No. W911NF-14-1-010. A.B. acknowledges support from the Ram\'on y Cajal program under RYC-2016-20066, Spanish MINECO project FIS2015-70856-P, and CAM PRICYT project QUITEMAD+ S2013/ICE-2801. S. C. B. acknowledges EPSRC project EP/M013243/1. We gratefully thank all members of the eQual (\textit{Encoded Qubit Alive}) consortium for their continuous input and useful discussions during several stages of this work.

\appendix

\section{\bf Trapped-ion quantum information processing (QIP):  microscopic QEC toolbox and error models}
\label{sec:appendix_A}

In this Appendix, we present a microscopic description of the QEC toolbox and the error models with details omitted in the main part of the text, focusing on $^{40}$Ca$^+$ optical qubits to store and manipulate the quantum information.

{\it (a) Coherent errors  for single-qubit rotations.--} We start by describing in detail the errors for single-qubit gates~\eqref{eq:global_rotations}-\eqref{eq:local_rotations}, which are driven by lasers tuned to the so-called carrier transition, or highly off-resonant lasers leading to an ac-Stark shift.

 \begin{figure*}[t]
	\begin{centering}
		\includegraphics[width=1.4\columnwidth]{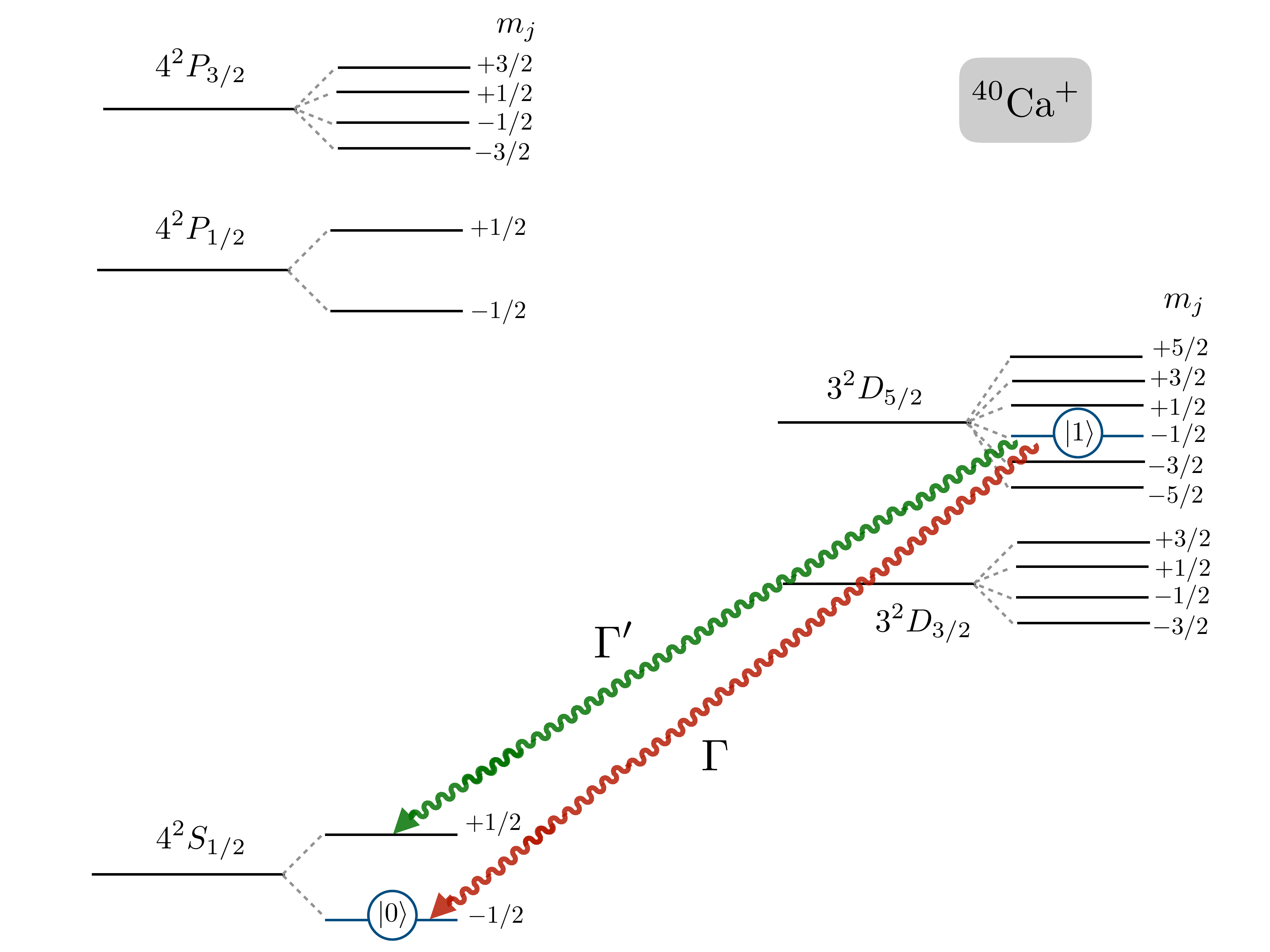}\\
		\caption{ \label{fig:damping_leakage} {\bf Amplitude damping and leakage for the $^{40}{\rm Ca}^+$ optical qubit:}  Level structure for the $^{40}{\rm Ca}^+$ optical qubit with $\ket{0}=\ket{S_{1/2},m_j=-1/2}$  and $\ket{1}=\ket{D_{5/2},m_j=-1/2}$. The finite lifetime of the metastable state is $T_1=1/\Gamma\approx 1.1\,$s, and there is a finite branching ratio $\Gamma'/\Gamma= 4/9$, leading to a finite probability that the spontaneous decay results in population of  the $S_{1/2},m_j=+1/2$ level. According to our computational model, $\Gamma$ represents the damping rate, while $\Gamma'$ represents the smaller leakage rate.}
	\end{centering}
\end{figure*}

For the carrier transition,  quantum gates~\eqref{eq:global_rotations} are implemented by coupling the lowest-lying $S$-state to a metastable $D$-state forming the optical qubit $\ket{0}=\ket{4S_{1/2},-1/2}$ $\ket{1}=\ket{3D_{5/2},-1/2}$, which are  labelled by the principal quantum number and various  orbital/spin angular momenta $\ket{nL_J^{2S+1},M_J}$) (see Fig.~\ref{fig:damping_leakage}) via resonant light at $729\,$nm. The coupling strength of the laser-ion interaction is given by the Rabi frequency $\Omega$, whereas the product between Rabi frequency $\Omega$ and  interaction time $\tau_{\theta}$ yields the pulse area $\theta = \Omega\cdot\tau_{\theta}$.
Note the proportionality between squared Rabi frequency and the laser  intensity $\Omega^2\propto I_0$. On a Bloch sphere picture, the pulse area corresponds to a rotation angle $\theta$ around a specific axis determined by the azimuthal angle $\phi$. The rotation axis is restricted to the equatorial plane and defined by adjusting the relative phase between laser and ion. Therefore, the carrier quantum gates~\eqref{eq:global_rotations} applied to a single qubit can be described as follows
\begin{equation}
\label{Eqn3:1QubitGatter}
U_{\rm R,\phi}(\theta) = \cos\frac{\theta}{2}\mathbb{I} + i\sin\frac{\theta}{2}\Big(\cos\phi\hat{X}+\sin\phi\hat{Y}\Big)\;.
\end{equation}
One can clearly see that laser intensity fluctuations will have an impact on the Rabi frequency $\Omega(t)$, and therefore induce an incorrect pulse area $\theta(t)$. Additionally, fluctuations in the laser phase $\phi(t)$ will lead to an error in the orientation of the rotation axis. The noisy gate should thus be obtained from the time-ordered exponential $U_{{\rm R,}\phi(t)}(\Omega(t))=\mathsf{T}\left\{\ee^{-\ii\int{\rm d}t\Omega(t)X_{\phi(t)}}\right\}$, which will depend on the particular dynamical pattern of fluctuations.  Assuming that the phase fluctuations occur on a much slower timescale than the gates, as is typically the case in experiments,  we can obtain an error model where phase and pulse-area fluctuations have clearly separated effects
\begin{equation}
\label{Eqn3:FehlerhafteGatter}
\begin{split}
U_{{\rm eff,}\phi(t)}(\theta(t))&= \cos\frac{\theta_{\text{eff}}(t)}{2}\mathbb{I} \\
&+ i\sin\frac{\theta_{\text{eff}}(t)}{2}\Big(\cos\phi\hat{X}_{\text{eff}}(t)+\sin\phi\hat{Y}_{\text{eff}}(t)\Big).
\end{split}
\end{equation}
Here, intensity fluctuations introduce a noisy rotation angle
\begin{equation}
\theta_\text{eff}(t) = \theta\cdot\sqrt{\frac{I(t)}{\langle I(t)\rangle}},
\end{equation}
while a fluctuating phase $\phi(t)=\phi+\delta \phi(t)$ between laser light and ion changes the orientation of the rotation axis from the ideal transformation due to a slower phase drift between consecutive gates, 
\begin{equation}
	\label{Eqn3:AchseEff}
	\begin{aligned}
	\hat{X}_\text{eff}(t) &= \cos\delta\phi(t)\hat{X} + \sin\delta\phi_{F}(t)\hat{Y}, \\
	\hat{Y}_\text{eff}(t) &= \cos\delta\phi(t)\hat{Y}
-\sin\delta\phi(t)\hat{X}. 	\end{aligned}
	\end{equation}
These equations are readily generalized to the multi-ion case where the carrier gates act globally on all ions residing in the same trap region~\eqref{noisy_carrier}. 

Regarding the local rotations~\eqref{eq:local_rotations} via ac-Stark shifts, we note that these two-photon processes will be largely insensitive to slow phase drifts. Therefore, they will only suffer from intensity fluctuations, which modify the pulse area and the rotation angle according to Eq.~\eqref{eq:noisy_local_rotations}.
 
The next ingredient in the noise model is to use a particular stochastic process for the intensity and phase fluctuations. In this work, we use the so-called Ornstein-Uhlenbeck random process $F(t)$~\cite{random_processes,ou_random_process}, which  evolves  under the following  Langevin equation
\begin{equation}
\frac{{\rm d}F(t)}{{\rm d}t}=-\frac{F(t)}{\tau_c}+\sqrt{c}\Gamma(t).
\end{equation}
Here, $c$  is the diffusion constant of the random process,  $\tau_c$ the correlation time, and  $\Gamma(t)$ is a Gaussian white noise with averages $\overline{ \Gamma(t)}=0$, $\overline{\Gamma(t)\Gamma(0)}=\delta(t)$, where $\delta(t)$ is the Dirac delta modelling the uncorrelated noise. This  stochastic differential equation can be integrated exactly yielding a Gaussian random process with autocorrelation function  $\overline{ F(t)F(0)}=\frac{c\tau_c}{2}\ee^{-t/\tau_c}$. The idea is that, by adjusting the model constants, we can numerically simulate  both the intensity and laser phase noise with their different characteristics using different numerically-generated random processes $\{F_1(t),F_2(t)\}\to \{\phi(t),\theta(t)\}$. 

 This type of noise modeling is well-suited for its implementation in a pure-state Monte Carlo formalism where parallelism is exploited to calculate averages over the stochastic time-dependent noise. Essentially, for each  idling time $t_{\rm I}$, one must average over $\ket{\psi(t+t_{\rm I})}=U_{\rm eff,\phi(t+t_{\rm I})}(\theta(t+t_{\rm I}))\ket{\psi(t)}$ for the different noise realizations. Numerically, one discretizes the time interval in steps of $\delta t=t_{\rm I}/N$, and computes
\beq
\label{unitary_noise}
\ket{\psi(t+t_{\rm I})}\approx  \prod_{m=1}^N U_{{\rm eff},\phi(t_m)}(\theta(t_m))\ket{\psi(t)}.
\eeq
Here, the values of the stochastic process are calculated by the update formula of the Ornstein-Uhlenbeck process, which is valid for any discretization $t_2=t_1+\delta t$
\begin{equation}
\label{noise}
F(t_2)=F(t_1)\ee^{-\frac{\delta t}{\tau_c}}+\big[\textstyle{\frac{c\tau_c}{2}}(1-\ee^{-\frac{2\delta t}{\tau_c}})\big]^{\frac{1}{2}}n,
\end{equation}
where $n$ is normal random variable of mean 0 and  variance 1. Different unitary time-evolutions~\eqref{unitary_noise} are calculated in parallel for different  values of $n$, which generate different samplings of the process $F(t)\in\{\theta(t),\delta\phi(t)\}$, and are incorporated in our full wave-function simulations.

For the numerical simulation of   the phase noise, we have considered $\delta t=1\,\mu$s, and $\tau_c=0.1\,$s such that we are close to the aforementioned Wiener model with $c=0.01$. For the intensity fluctuations, we take a much shorter correlation time $\tau_c=T_2/1000$. These parameters are 
are chosen such that the fidelity of the various gates coincides with the values  listed in  table~\ref{tab:summary_toolbox}.

\vspace{1ex}
{\it (b) Microscopic errors  for the entangling gates}.-- Let us now discuss the microscopic error model for the entangling gates~\eqref{eq:MS_gate}. We consider the MS gates mediated by the longitudinal center-of-mass (CoM) mode of a crystal of two $^{40}{\rm Ca}^+$ ions. As a starting point, we will consider an MS gate that suffers from {\it motional errors} from both the CoM and stretch modes (i.e.~residual spin-phonon entanglement and Debye-Waller fluctuations of the Rabi frequency), and {\it collective dephasing} (i.e.~decoherence due to fluctuations of global magnetic fields). These Debye-Waller factors are caused by the thermal population of the vibrational modes, both bus and spectator modes~\cite{PhysRevA.62.022311}.  This model will be the starting point that can be improved by incorporating the effects of the off-resonant carrier, and  fluctuating laser intensities/phases.

 \begin{figure*}[t]
 \begin{centering}
  \includegraphics[width=1.7\columnwidth]{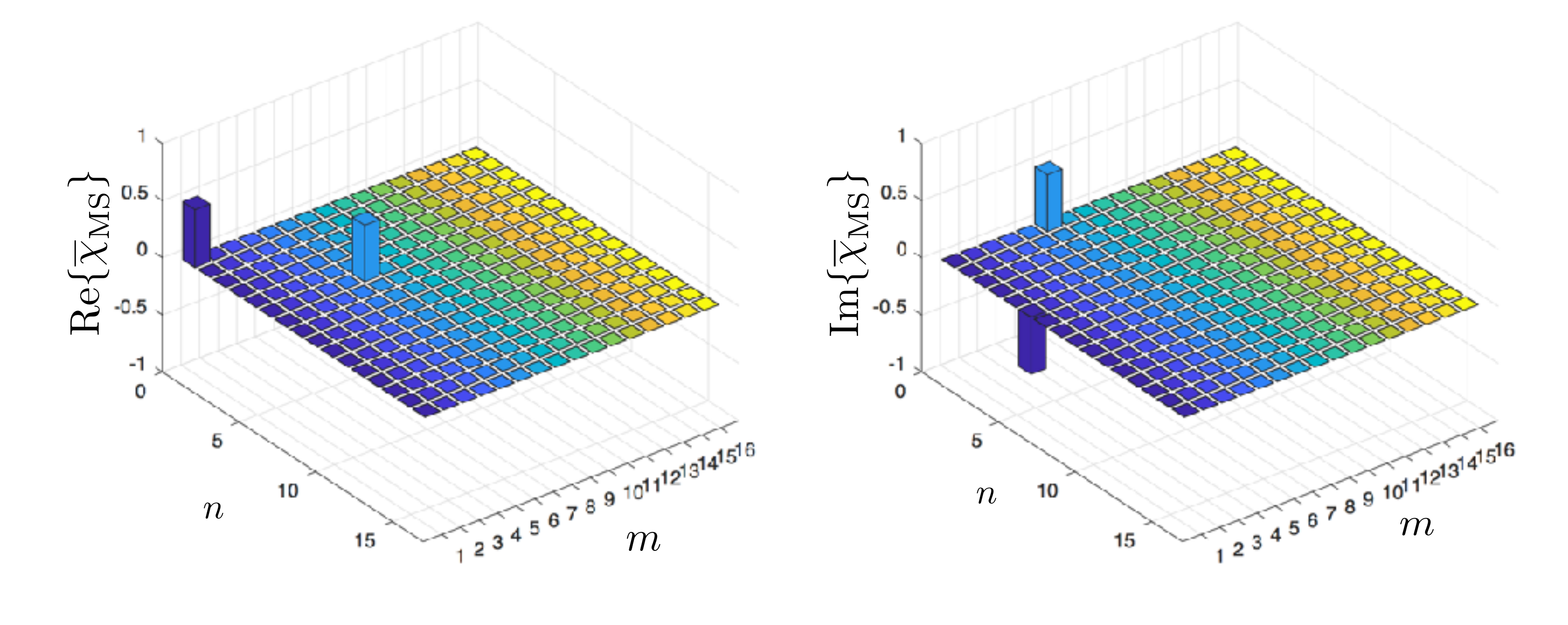}\\
\caption{ \label{fig:chi_MS} {\bf Average $\chi$ matrix for a noisy MS gate:} Longitudinal MS gate for two $^{40}{\rm Ca}^+$ ions, where the axial trap frequency is $\omega_z/2\pi=0.975\,$MHz. We consider that  the longitudinal modes are laser cooled to $\bar{n}_{\rm CoM}=0.1$, and  $\bar{n}_{\rm stretch}=0.016$, and the CoM mode is exploited to mediate a one-loop MS gate with a laser Rabi frequency of $\Omega_{\rm L}/\omega_z=0.1$. We model the collective dephasing with a Markovian Ornstein-Uhlenbeck process associated to $T_2=0.2s$.  }
\end{centering}
\end{figure*}

At this level, the time-evolution of the two qubits can be formally written as follows
\beq
\label{eq:MS_gate_ev_app}
\rho(t_{\rm g})={\rm Tr}_{\rm ph}\left(
U_{\rm g}\rho_0U^\dagger_{\rm g}
\right),\hspace{1ex} U_{\rm g}=\ee^{-\ii t_{\rm g}H_0}\mathsf{T}\left\{\ee^{-\ii\int_0^{t_{\rm g}}{\rm d}t' H_{\rm c}(t')}\right\},
\eeq
where $t_{\rm g}$ is the gate time, and $H_0=\sum_i\frac{\omega_0}{2}\sigma_i^z+\sum_n\omega_na^\dagger_na_n$ is the Hamiltonian for the uncoupled qubits and longitudinal phonons of frequency $\omega_0$ and $\{\omega_n\}_{n=1}^N$, respectively. The qubit-phonon coupling in the MS scheme with the above sources of noise can be written as
\beq
\label{eq:noisy_state_dep_force}
H_{\rm c}(t)=\sum_i\frac{ F(t)}{2}\sigma_i^z+\sum_{i,n}\hat{\frak{F}_{in}}\sigma_i^{\phi}a_n\ee^{-\ii\omega_nt}\cos(\delta t)+{\rm H.c.},
\eeq
where we have used an Ornstein-Uhlenbeck random process $F(t)$, similar to the ones used to model laser fluctuations in single-qubit gates, to model the  dephasing during the MS gate. Additionally, the second term represents a state-dependent dipole force proportional to $\sigma_i^\phi=\cos\phi\sigma_i^x-\sin\phi \sigma_i^y$, where $\phi$ is the common   phase of  a pair of laser beams tuned to the red and blue motional sidebands of the ions with  opposite detunings (i.e. $\omega_{\rm L}=\omega_0\pm \delta\approx\omega_0\pm\omega_z$, where $\omega_z$ is the axial trap frequency that coincides with the lowest-frequency center-of-mass-mode $\omega_1=\omega_z$). We note that the  strength of the  force $\hat{\frak{F}}_{jn}=\frak{F}_{jn}(1-\half\eta_na_n^\dagger a_n)$ is proportional to the laser intensity $\frak{F}_{jn}\propto I_{\rm L}$, which can also be fluctuating, but also depends on the phonons (i.e. Debye-Waller effect) where  $\eta_n$ is the Lamb-Dicke parameter.

This equation is precisely the starting point to calculate the quantum state fidelity of a maximally entangled state as generated by a fully-entangling MS gate, and extract the noise parameters of the effective gate error model of Ref.~\cite{eQual_1qubit}. In this part of the Appendix, we describe  a different approach that relies on the full numerical simulation of the unitary for time evolution, from which one can obtain a quantum channel for the reduced density matrix describing the trapped-ion qubits that is not based on a single number like the aforementioned state fidelity~\cite{eQual_1qubit}. Evaluating the above time evolution~\eqref{eq:MS_gate_ev} using  a pure-state  formalism can only be achieved for a perfectly groundstate-cooled crystal. In more realistic situations, the vibrations are in a low-excitation thermal state after  laser cooling, and one has to treat the full density-matrix evolution. Since we ultimately want to describe the noisy MS gate within the Monte Carlo pure-state formalism, we need to  find  a  quantum channel that describes the reduced dynamics of the two qubits by the MS gate
\beq
\label{eq:MS_channel}
\rho(t_{\rm g})=\sum_np_nK_n\rho_0K_n^{\dagger}, \hspace{3ex}\sum_np_nK_n^\dagger K_n=\mathbb{I}.
\eeq
Here,  $K_n$ are the  two-qubit Kraus operators to be found, and
$p_n$ are their corresponding probabilities. In former studies,  the  noisy  MS gate $\rho(t_{\rm g})=\epsilon_{\rm MS}(U_{\rm MS}\rho_0U_{\rm MS}^\dagger)$ was described as  the ideal gate  $U_{\rm MS}=(\mathbb{I}-\ii \sigma_{i_1}^\phi \sigma_{i_2}^\phi)/\sqrt{2}$ followed by an idealized depolarising channel~\cite{nielsen-book}, $\epsilon_{\rm MS}(\bullet)$, which depends on the number of qubits involved in the gate~\cite{eQual_1qubit}. In this work, we improve upon this model, and  describe the  noisy  MS gate directly in terms of the exact  set of Kraus operators with certain probabilities that can be numerically computed~\eqref{eq:MS_channel}. We note that both descriptions can be readily incorporated in a pure-state Monte Carlo evolution.

We  obtain the desired quantum channel~\eqref{eq:MS_channel} numerically by truncating the vibrational Hilbert-space, calculating the full time-evolution operator $U_{\rm g}$~\eqref{eq:MS_gate_ev},  and performing quantum process tomography to determine the corresponding $\{p_n,K_n\}$~\cite{process_tomography}. For two qubits, a generic quantum channel can be expressed in the so-called $\chi$-matrix representation $\rho(t_{\rm g})=\sum_{n,m}\chi_{n,m}E_n\rho_0E_m^\dagger$, where the set of 16 operators $\{E_n\}$ is obtained from all the possible tensor products of $\{\mathbb{I},\sigma^x,\ii\sigma^y,\sigma^z\}$.  The $\chi$ matrix can be extracted from the action of the microscopic time evolution ${\rm Tr}_{\rm ph}\left(U_{\rm g}\rho_jU^\dagger_{\rm g}\right)$ and that of the quantum-channel operators $E_n\rho_jE_m^\dagger$ on a set of initial states  $\{\rho_j\}$ that forms a basis of the space of operators. For the two-qubit case, we follow a compact recipe~\cite{process_tomography} to extract the $\chi$ matrix for the noisy MS gate, which can be diagonalised
\beq
\chi=\sum_np_n\ket{v_n}\bra{v_n},\hspace{2ex} K_n=\sum_m \langle e_m|v_n\rangle E_m,
\eeq
where  $\{p_n,\ket{v_n}\}$ are the eigenvalues and eigenvectors of the $\chi$ matrix, and $\{\ket{e_m}\}$ are the Cartesian unit vectors of a 16-dimensional vector space.

According to this discussion, we can reconstruct the quantum channel for the MS gate~\eqref{eq:MS_channel} by simply diagonalising the $\chi_{\rm MS}$ matrix associated to the microscopic evolution~\eqref{eq:MS_gate_ev} for each realisation of the random process that models the collective dephasing noise. After averaging over $N_{\rm s}$ random samplings of the noise, we can build $\bar{\chi}_{\rm MS}$ and extract the average Kraus operators for $\overline{K}_n$ with probabilities $\overline{p}_n$. In Fig.~\ref{fig:chi_MS}, we represent the real and imaginary parts of the average $\chi$-matrix representation of a noisy MS gate that generates a maximally-entangled state with average error $\bar{\epsilon}_{\rm MS}=8.44\cdot 10^{-4}$ in $t_{\rm g}=72\mu$s.  For these values, we obtain the following probabilities $\overline{p}_{n}\in\{0.999, 0.0005, 0.0002, 0.0001, \cdots\}$, where the dots represent probabilities below $10^{-8}$. Essentially, $\overline{K}_1$ is very similar to the ideal gate $U_{\rm MS}=(\mathbb{I}-\ii \sigma_{i_1}^\phi \sigma_{i_2}^\phi)/\sqrt{2}$, whereas the other Kraus operators $\overline{K}_2,\overline{K}_3,\overline{K}_4$ represent single and two-qubit errors in the basis of the MS gate. As anticipated above, this more realistic, microscopically derived error channel differs substantially from the depolarising channel used in~\cite{eQual_1qubit}.

 We note that this approach can be easily incorporated in the Monte Carlo numerical simulations of previous sections. First, we want to use the set $\{\overline{p}_n,\overline{K}_n, \forall n: \overline{p}_n>p_{\rm trunc}\}$ to approximate the microscopic channel~\eqref{eq:MS_channel}. For a pure-state Monte Carlo evolution, we  need to generate random numbers $r \in[0,1]$ and apply the numerically generated $\overline{K}_n$ if $r$ falls in the respective probability interval, $\sum_{k}^{n-1} \overline{p}_{k} \leq r < \sum_{k}^{n} \overline{p}_{k}$, where $p_0=0$. In this way, one randomly samples over all the relevant Kraus operators, such that the stochastic average yields the desired evolution of the noisy MS gate.

From the initial  experience gained with this microscopic numerical modeling, we can now account for another important source of errors that is not considered in Eq.~\eqref{eq:noisy_state_dep_force}. Although the dephasing and thermal noise can be the leading source of MS gate error in situations where the $T_2$ time is much shorter, or where the vibrations are not cooled to sufficiently small phonon occupation numbers (e.g. see some of the QEC performance of~\cite{{eQual_1qubit}}, where sympathetic re-cooling prior to the MS gates was not exploited), for the current regime of parameters, an off-resonant carrier term that acts in an orthogonal basis with respect to the state-dependent force can actually be the leading source of error.

Rather than treating this term  in perturbation theory, which underlies the analysis of~\cite{{eQual_1qubit}}, and the approach of  publications~\cite{PhysRevA.62.022311, molmer-prl-82-1835}, we use the formalism developed by C.F. Roos
~\cite{roos_gates} to take this term into account. By moving into a rotating frame with respect to the off-resonant carrier, the state-dependent dipole force in Eq.~\eqref{eq:noisy_state_dep_force} must be changed into
\beq
\label{eq:noisy_state_dep_force_off_carrier}
\begin{split}
H_{\rm c}(t)\approx &\sum_{i,n}{\frak{F}}_{in}(J_0+J_2)(\sigma_i^{x}\cos\Psi+\sigma_i^{z}\sin\Psi)a_n\ee^{\ii(\zeta-\omega_nt)}\cos(\delta t)\\
&+{\rm H.c.},
\end{split}
\eeq
where we have introduced the phase difference $\zeta$ between the laser beams driving the blue and red sidebands, and $\Psi=2\Omega_{\rm L}\sin\zeta/\omega_{\rm L}$ leads to an intensity-dependent rotation angle over the basis of the MS gate, as can be seen by comparing Eq.~\eqref{eq:noisy_state_dep_force_off_carrier} to Eq.~\eqref{eq:noisy_state_dep_force}. In addition, we have introduced the first-class Bessel functions $J_0=J_0(2\Omega_{\rm L}/\omega_{\rm L})$, and $J_2=J_2(2\Omega_{\rm L}/\omega_{\rm L})$. We note that for the numerical simulation, the additional magnetic-field noise of Eq.~\eqref{eq:noisy_state_dep_force} is also included, as well as small additional  contributions to Eq.~\eqref{eq:noisy_state_dep_force_off_carrier} that stem from the Debye-Waller factors mentioned above, and  effective spin-spin interactions that appear as one moves onto the aforementioned rotating frame (see Ref.~\cite{roos_gates} for details on such small qubit-qubit couplings, which also depend on $J_1=J_1(2\Omega_{\rm L}/\omega_{\rm L})$). In addition to the dephasing and motional errors also accounted for in Eq.~\eqref{eq:noisy_state_dep_force}, this new formulation~\eqref{eq:noisy_state_dep_force_off_carrier} allows us to account for the effects of the off-resonant carrier. Even if the effect of a finite $\zeta$ and $\psi$ can be partially overcome by an adiabatic switching of the forces, together with a refocusing pulse shaping that inverts the sign of the state-dependent force at the middle of the gate, slow fluctuations in the intensity (and thus on $\zeta$) yield a residual error that can be a leading source of infidelity of the MS gate. In order to capture these effects in the effective noise model, we repeat the above procedure of process tomography, but this time  using Eq.~\eqref{eq:noisy_state_dep_force_off_carrier}, together with the additional terms, in the  numerical simulation. In these numerics, we set the parameters for a perfect entangling gates at $\zeta=0$, and then modify its value to account for possible drifts and errors that are in accordance to Table~\ref{tab:summary_toolbox}. We again perform process tomography, and extract a set  $\{\overline{p}_n,\overline{K}_n, \forall n: \overline{p}_n>p_{\rm trunc}\}$, which is directly fed into the full wave-function numerical simulations.

\vspace{1ex}
{\it (c) Amplitude damping and qubit leakage}.-- Let us now discuss the details of environmental spontaneous decay which, in addition to amplitude damping, can also populate the groundstate Zeeman sublevel $S_{1/2}(m_j=+1/2)$ lying outside of the computational subspace (i.e. {\it leakage}), as depicted in Fig.~\ref{fig:damping_leakage}. In this part of the Appendix, we give some of the details of  the circuit model for the simulation of this process that were omitted in the main text.

Once again, let us start by considering the circuit model of amplitude damping (see the left panel of  Fig.~\ref{damping_leakage_circuit}{\bf (a)}, and chapter 8 of~\cite{nielsen-book}). In this figure, the controlled rotation of angle $\theta_{\rm d}$ can be expressed as $U_{CR}= \half(1-Z_d)\ee^{-\ii\frac{\pi}{4}Y_a}+\half(1+Z_a)\mathbb{I}_a$, while the CNOT  is $U_{CNOT}=\half(1-Z_a)\ee^{-\ii\frac{\pi}{2}X_d}+\half(1+Z_a)\mathbb{I}_d$, where we label the data qubit with $d$ and the ancillary one with $a$. After the measurement, the state of the data qubit will be
\beq
\rho_{\rm f}={\rm Tr}_a\{P_{a,0}\rho' P_{a,0}+P_{a,1}\rho'P_{a,1}\}
\eeq
where $\rho'=U_{CNOT}U_{CR}(\rho\otimes\ket{0}_a\bra{0}_a)U _{CR}^\dagger U_{CNOT}^{\dagger}$ for an arbitrary state of the data qubit $\rho$, and where we have introduced the ancilla projectors onto the  computational basis states $P_{a,0/1}$. Accordingly, we only need to look at the diagonal elements  of the transformed density matrix 
\beq
\begin{split}
\rho'_{11}&=\sin^2(\theta_{\rm d}/2)\rho_{11}\ket{1}_d\bra{1}_d,\\
\rho'_{00}&=\rho_{00}\ket{0}_d\bra{0}_d+\cos(\theta_{\rm d}/2) \bigl (\rho_{10}\ket{1}_d\bra{0}_d+\rho_{01}\ket{0}_d\bra{1}_d \bigl )\\
&+\cos^2(\theta_{\rm d}/2)\rho_{11}\ket{1}_d\bra{1}_d.
\end{split}
\eeq
One can easily check that this evolution is equivalent to that of the amplitude-damping channel $\rho_{\rm f}=\epsilon_{\rm d}(\rho):=L_0\rho L_0^{\dagger}+L_1\rho L_1^{\dagger}, $ with the following Kraus operators
\beq
\begin{split}
L_0 &= \ket{0}_d\bra{0}_d+\sqrt{1-p_{\rm d}}\ket{1}_d\bra{1}_d, \\
L_1 &= \sqrt{p_{\rm d}}\ket{0}_d\bra{1}_d,
\end{split}
\eeq
where one finds that the  angle of the controlled rotation must be fixed by the decay parameter $p_{\rm d}=\sin^2(\theta_{\rm d}/2)$, where $p_{\rm d}= 1- \exp(-\Gamma t)$. We note that, at the level of the reduced density matrix of the data qubit, the circuit of the left  of the upper panel of Fig.~\ref{damping_leakage_circuit} is equivalent to that in the right, where the classical information of the measurement is used to apply a conditional $X$ gate on the data qubit with probability $p_{\rm d}=\sin^2(\theta_{\rm d}/2)$ (i.e. only when the measurement result indicates that the ancillary qubit was in $\ket{1}_a$).

 \begin{figure*}[t]
 \begin{centering}
  \includegraphics[width=1.4\columnwidth]{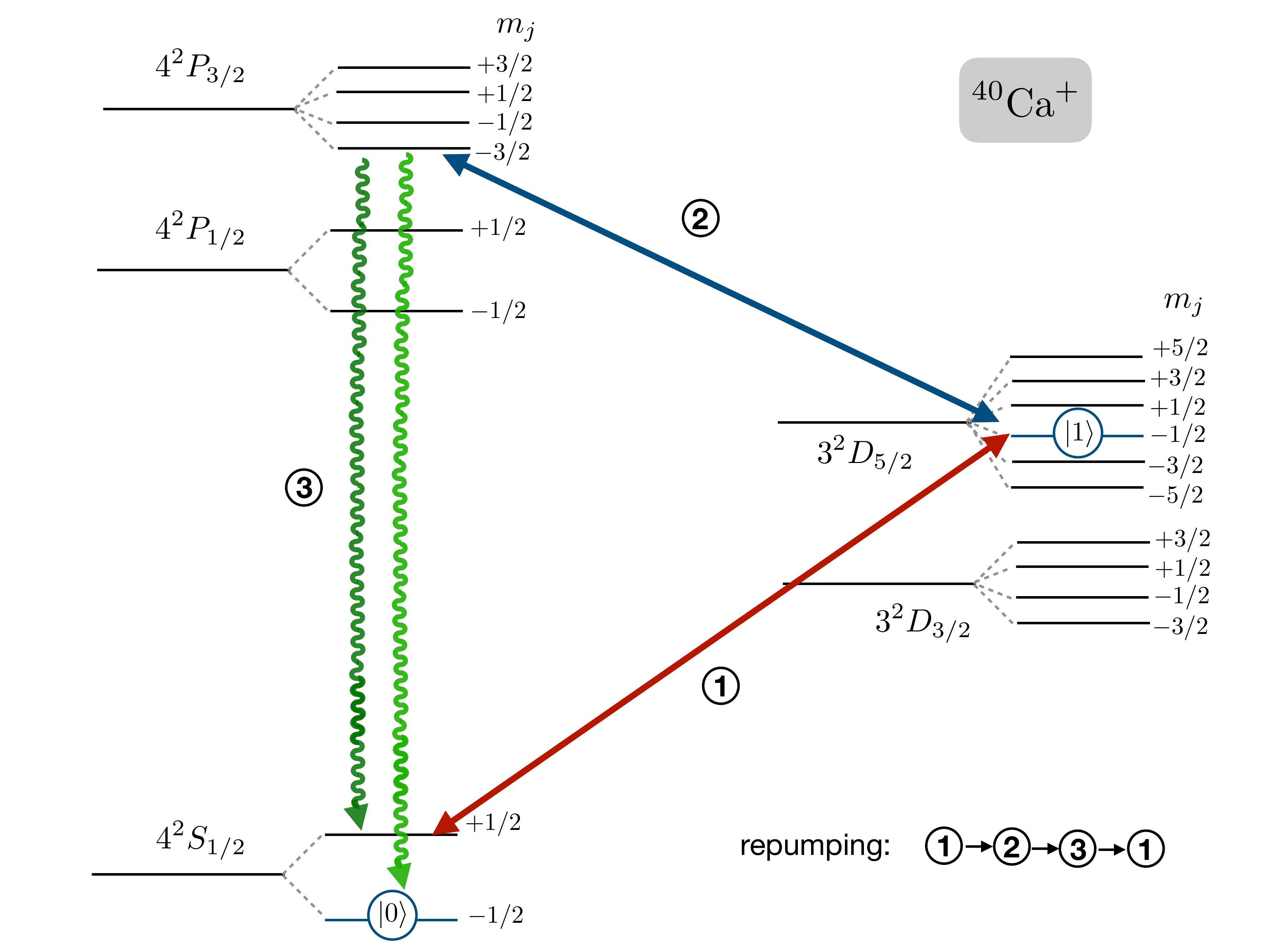}\\
\caption{ \label{fig:damping_leakage_repumping} {\bf Repumping scheme  for the   $^{40}{\rm Ca}^+$ optical qubit}:~The repumping cycle consists of (1) applying a $\pi$-pulse that brings the population of the leaked level to the metastable state, or otherwise hides the qubit in the $\ket{1}$ state into the leaked level. Then, (2) one applies a laser driving the dipole-allowed transition to the excited $P_{3/2}$ level, such that the population of the leaked state will (3) spontaneously decay into the groundstate manifold. By finally applying (1) another $\pi-$pulse, the population is  re-pumped back into the computational subspace, albeit losing coherences and affecting the information stored in the qubit.}
\end{centering}
\end{figure*}

In order to simulate the simultaneous amplitude damping and leakage of Fig.~\ref{fig:damping_leakage}, we can build on this philosophy and, as argued in the main text, use an additional classical bit to store  the information about the leaked level (i.e.~$\ell=1$ if the qubit has not leaked into the $S_{1/2},m_j=+1/2$ level, and $\ell=0$ if the qubit has indeed leaked). The leakage with rate $\Gamma'$, which can be simulated by means of an ancilla qubit $a_1$ that is subject to a controlled rotation $U_{CR}=(\half(1-Z_d)\ee^{-\ii\frac{\pi}{4}Y_{a_1}}+\half(1+Z_d)\mathbb{I}_{a_1})\delta_{\ell,1}+\delta_{\ell,0}\mathbb{I}$ that is only applied when the qubit is still not leaked (i.e.~conditional on the classical bit being $\ell=1$, as depicted in the lower panel of Fig.~\ref{damping_leakage_circuit}). After measuring the ancillary qubit, one applies a conditional classical  NOT operation on the classical bit, which will turn it into the leaked state $\ell=0$ with probability $\sin^2(\theta_\ell/2)=p_{\ell}(t)=\Gamma' \left(1-\ee^{-(\Gamma+\Gamma')t}\right)/(\Gamma+\Gamma')$, which is obtained by solving the corresponding master equation for the population $p_{\ell}(t)$ explicitly. 

After that, a different controlled rotation $U_{CR}'=(\half(1-Z_d)\ee^{-\ii\frac{\pi}{4}Y_{a_2}}+\half(1+Z_d)\mathbb{I}_{a_2})\delta_{\ell,1}+\delta_{\ell,0}\mathbb{I}$ is applied to a fresh ancillary qubit $a_2$, which is conditioned on the classical bit being $\ell=1$ with probability $(1-p_{\ell}(t))$. This indicated that the qubit is still not leaked, and can thus  decay into the $\ket{0}=\ket{S_{1/2},m_j=-1/2}$ state. In that case, the ancilla qubit $a_2$ is rotated by an angle $\theta_{\rm d}$ when the data qubit lies in the computational $\ket{1}=\ket{D_{5/2},m_j=-1/2}$ state, such that it can indeed decay by spontaneous emission. The corresponding amplitude damping is simulated by the final $X$ gate conditioned on the measurement result. Accordingly, the probability of the damping channel is $p_{\rm d}=(1-p_{\ell})\sin^2(\theta_{\rm d})$, which must be equal to the physical value $p_{\rm d}(t)=\Gamma\left(1-\ee^{-(\Gamma+\Gamma')t}\right)/(\Gamma+\Gamma'\ee^{-(\Gamma+\Gamma')t})$, and gives the condition to set the rotation angle  to the correct value $\sin^2(\theta_{\rm d}/2)=p_{\rm d}(t)$. As a summary, the rotation angles obtained are those of Eq.~\eqref{eq:decay_angles} of the main text.

This circuit is equivalent to the amplitude damping and leakage
channel $\rho_{\rm f}=\epsilon_{\rm d,l}(\rho):=L_0\rho L_0^{\dagger}+L_1\rho L_1^{\dagger}+L_2\rho L_2^{\dagger}, $ with the following Kraus operators
\beq
\begin{split}
L_0 &= \ket{0}_d\bra{0}_d+\ket{\ell}_d\bra{\ell}_d+\sqrt{1-p_{\rm d}-p_{\ell}}\ket{1}_d\bra{1}_d \\
L_1 &= \sqrt{p_{\rm d}}\ket{0}_d\bra{1}_d \\
L_2 &= \sqrt{p_{\ell}}\ket{\ell}_d\bra{1}_d.
\end{split}
\eeq

Let us now describe the microscopic details of the re-pumping scheme, which consists on steps (1)-(2)-(3)-(1) (see Fig.~\ref{fig:damping_leakage_repumping}). Here, (1) corresponds to a $\pi$-pulse between  $\ket{D_{5/2},m_j=-1/2}$ and the leaked state $\ket{S_{1/2},m_j=1/2}$. If the state of the qubit has not leaked, the quantum information is protected as the remaining operations (2)-(3) do not take place, and the final (1) $\pi$-pulse  brings the population back to $\ket{D_{5/2},m_j=-1/2}$, yielding the initial un-leaked qubit state. On the other hand, if the qubit had indeed leaked into the $\ket{S_{1/2},m_j=1/2}$ level, (2) the lasers tuned to the dipole-allowed transition will bring this population up to an excited P level, which (3) will decay very fast into the $S_{1/2}$ groundstate manifold. At this point, the leakage has become a sort of amplitude damping, while the original coherences of the qubit state hidden in the groundstate manifold are still present. Then, a final (1) $\pi$-pulse between brings the qubit back to the computational space  $\ket{S_{1/2},m_j=-1/2}, \ket{D_{5/2},m_j=-1/2}$, such that subsequent rounds of QEC can project it back onto the stabiliser subspace.

 Clearly, this repumping scheme will not be perfect, since the $\pi$-pulses will be faulty, and there might be branching to other levels as well. As a first error model, we consider that $\epsilon$ is related to the infidelity of the $\pi$-pulses in the repumping, and other possible imperfections. At the level of our circuit model with the classical bit, we fail to repump with probability $\epsilon^2$, such that $\ell=0$ remains in the leaked bit. On the other hand, with probability $1-\epsilon^2$, we re-pump into a mixed state in the computational basis.

\section{\bf Microscopic QEC schedules}
\label{sec:appendix_B}

As advanced in the main text, the different circuits for QEC correction, which are expressed in terms of the trapped-ion native set of gates Figs.~\ref{Fig:flag_readout_ions},~\ref{Fig:DVA_readout_ms} and~\ref{Fig:error_correction_encoding}, must be translated into detailed microscopic schedules for their implementation in the QCCD segmented trap. This schedules consist on specific sequences of elementary operations, combining quantum gates and crystal reconfiguration operations (see Table~\ref{tab:summary_toolbox}), which allow to implement  the desired approaches to QEC, and the Alice-Igor-Bob protocol to assess the integrity of the encoded memory.

\begin{figure*}[t]
 \begin{centering}
  \includegraphics[width=2\columnwidth]{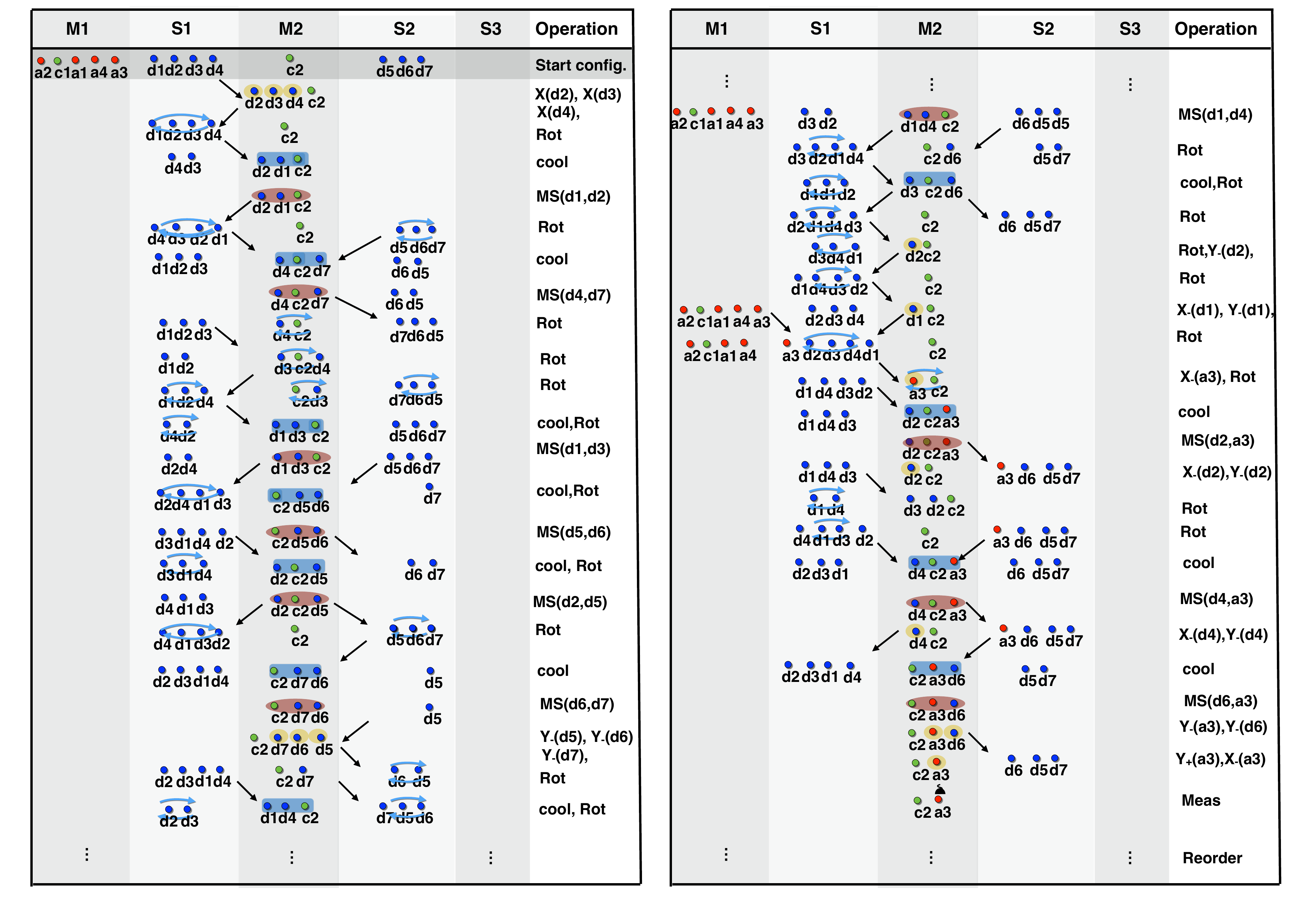}\\
  \caption{\label{Fig:schedule_encoding} {\bf Schedule for the efficient encoding in the 7-qubit color code:} Sequence of  trapped-ion operations  that must be applied to realise the  encoding circuit for the logical $\ket{0}_{\rm L}$ in Fig.~\ref{Fig:efficient_encoding_ft_ms}. The data $\{d1,\cdots,d7\}$, ancillary $\{a1,\cdots, a4\}$, and cooling $\{c1,c2\}$ ions are represented with different colours following the convention of Fig.~\ref{Fig:trap_gate_set}. These ions are arranged in different zones of a single arm of the trap, following the labelling of Fig.~\ref{Fig:trap_gate_set}. We distribute these ions among the different columns of the table, reserving the rightmost column to list the operations that bring one configuration to the following one, where time  flows downwards in the table. The split/merge and shuttle operations are represented by arrows, and can be deduced from consecutive ion configurations.The remaining operations have labels in the rightmost column. In particular, $X_{\pm}(i)=X_i(\pm\pi/2)$, $Y_{\pm}(i)=Y_i(\pm\pi/2)$ are local rotations obtained from Eqs.~\eqref{eq:global_rotations}-\eqref{eq:local_rotations} by spin-echo pulses~\cite{eQual_1qubit}. These rotations act on the qubit $i$, and are depicted by yellow ellipses. The entangling gates are labelled as ${ MS}(i,j)$ corresponding to Eq.~\eqref{eq:MS_gate} for $\phi=0$ and $\theta=\pi$. This MS gate acts on the ion pair $i,j$, previously isolated by crystal reconfigurations, and is depicted by a a pink ellipse over the ions involved in the gate. Finally, $Rot$ stands for rotation of the crystal signalled by blue arrows, $Cool$ stands for sympathetic cooling of the crystal depicted with a blue ellipse, $Meas$ stands for the fluorescence measurement of the ancillary ion depicted with a semi-circular detector above the ion, and  $Reorder$ stands for crystal reconfigurations that bring the ions back to the original positions. Note that idle ions that do not take place in the set of operations are only drawn once in their corresponding column, until they are used at later stages, when they are explicitly drawn again.}
\end{centering}
\end{figure*}

Various microscopic schedules, regarding the cat-based approach to QEC, have been described in~\cite{eQual_1qubit}. Likewise, the trapped-ion set of microscopic instruction to perform flag-based QEC are contained in~\cite{latt_surgery}. In this appendix, we present the  missing microscopic schedule to implement a realistic Alice-Igor-Bob protocol, i.e. with imperfect encoding and decoding. Leet us focus on the FT encoding of the logical $\ket{0}_{\rm L}$ state using our toolbox  $\mathsf{(o1)-(o11)}$ for the mixed-species ion QCCD.  W start by considering the efficient encoding using the layout of ions in the DiVicenzo-Aliferis FT readout~\cite{eQual_1qubit}, which required 7 data qubits and 4 ancillary qubits for the  syndrome readout, both of which belong to the same atomic species. Additionally, this scheme exploited 2 cooling ions of a different species/isotope for sympathetic re-cooling of the ion crystal. The arrangement of these ions within the central region of the segmented trap is depicted in the inset of  Fig.~\ref{Fig:schedule_encoding}. Let us emphasize, however, that the schedule to be presented below only requires minor modifications to be adapted to other QEC schemes, such as the flag-based QEC, which requires a different configuration with only  7 data qubits, 2 ancillary qubits, and 1 cooling ions~\cite{latt_surgery}.

In Fig.~\ref{Fig:schedule_encoding}, we represent the initial distribution of the  13 ions  among the different zones (see the first line) of a single arm of the segmented trap (see Fig.~\ref{Fig:trap_gate_set}). The idea is that all the two-qubit MS gates, with the required sympathetic cooling, are to be applied in the M1 region by bringing the corresponding pairs of ions sequentially according to the order of the circuits in Fig.~\ref{Fig:efficient_encoding_ft_ms}. The microscopic schedule of Fig.~\ref{Fig:schedule_encoding} focuses on the encoding of the $\ket{0}_{\rm L}$ state, although that of $\ket{+}_{\rm L}$ has also been worked out and is being used in the numerical simulations. The subsequent lines of this figure represent a different step of the microscopic schedule, and the columns describe the particular ion occupation of each trap zone during such step. We use similar  conventions as in~\cite{eQual_1qubit,latt_surgery}, the operations to be performed  are listed in the right-most column,  and we also use straight black arrows to depict crystal splitting, shuttling, and merging; and curved arrows to denote crystal rotations (see the caption for further details). This microscopic schedule is translated into a quantum channel representation, which is then simulated numerically in combination with the QEC protocol to test for the beneficial role of Igor with imperfect Alice and Bob performance.

\end{document}